\documentclass[manuscript]{emulateapj}
\usepackage{natbib}
\usepackage{graphicx}
\usepackage{txfonts}

\makeatletter

\newcommand{\msun}{{$M_{\odot}~$}}
\newcommand{\msunp}{{$M_{\odot}$}}
\newcommand{\ser}{S\'ersic }

\newcommand{\mlg}{\mbox{$M_{\rm dyn}/L_{\rm{g}}$} }
\newcommand{\mlk}{\mbox{$M_{\rm dyn}/L_{\rm{K}}$} }
\newcommand{\gz}{\mbox{$(g-z)_{\rm rest-frame}$} }
\newcommand{\gk}{\mbox{$(g-K)_{\rm rest-frame}$} }

\slugcomment{Accepted for publication in ApJ}
\shorttitle{The Relation between Dynamical Mass-to-Light ratio and Color for Massive Quiescent Galaxies}
\shortauthors{Van de Sande et al.}

\begin{document}

\title{The Relation between Dynamical Mass-to-Light Ratio and Color for Massive Quiescent 
Galaxies out to $\lowercase{z}\sim2$ and Comparison with Stellar Population Synthesis Models}

\author{Jesse van de Sande\altaffilmark{1},
Mariska Kriek\altaffilmark{2},
Marijn Franx\altaffilmark{1}, 
Rachel Bezanson\altaffilmark{3}, 
Pieter G. van Dokkum\altaffilmark{4}
}

\altaffiltext{1}{Leiden Observatory, Leiden University, P.O.\ Box 
9513, 2300 RA Leiden, The Netherlands.}

\altaffiltext{2}{Astronomy Department, University of California at 
Berkeley, Berkeley, CA 94720, USA}

\altaffiltext{3}{Steward Observatory, University of Arizona, Tucson, AZ 85721, USA}

\altaffiltext{4}{Department of Astronomy, Yale University, P.O. \ Box
208101, New Haven, CT 06520-8101, USA.}


\begin{abstract}
\noindent We explore the relation between the dynamical mass-to-light ratio
($M/L$) and rest-frame color of massive quiescent galaxies out to
$z\sim2$. We use a galaxy sample with measured stellar velocity
dispersions in combination with \textit{Hubble Space Telescope} and
ground-based multi-band photometry. Our sample spans a large
range in $\log M_{\rm{dyn}}/L_{\rm{g}}$ (of 1.6~dex) and
$\log~M_{\rm{dyn}}/L_{\rm{K}}$ (of 1.3~dex). There is a strong,
approximately linear correlation between the $M/L$ for different
wavebands and rest-frame color. The root-mean-scatter scatter in
$\log~M_{\rm{dyn}}/L$ residuals implies that it is possible to
estimate the $M/L$ with an accuracy of $\sim0.25$~dex from a single
rest-frame optical color. Stellar population synthesis (SPS) models
with a  Salpeter stellar initial mass function (IMF) can not
simultaneously match $M_{\rm{dyn}}/L_{\rm{g}}$
vs. $(g-z)_{\rm{rest-frame}}$ and $M_{\rm{dyn}}/L_{\rm{K}}$
vs. $(g-K)_{\rm{rest-frame}}$. By changing the slope of the IMF we are
still unable to explain the M/L of the bluest and reddest galaxies. We
find that an IMF with a slope between $\alpha=2.35$ and $\alpha=1.35$
provides the best match. We also explore a broken IMF with a Salpeter
slope at $M<1M_{\odot}$ and $M>4M_{\odot}$ and a slope $\alpha$ in
the intermediate region. The data favor a slope of $\alpha=1.35$ over
$\alpha=2.35$. Nonetheless, our results show that variations between
different SPS models are comparable to the IMF variations. In our
analysis we assume that the variation in $M/L$ and color is driven by
differences in age, and that other contributions (e.g., metallicity
evolution, dark matter) are small. These assumptions may be an
important source of uncertainty as galaxies evolve in more complex
ways.
\end{abstract}


\keywords{galaxies: evolution --- galaxies: formation --- galaxies: kinematics and dynamics --- galaxies: stellar content --- galaxies: structure}

%

\section{Introduction} 

For a good understanding of galaxy evolution, accurate stellar mass
estimates are crucial (for a recent review see
\citealt{courteau2013}). Nearly all galaxy properties, among which
structure, star formation activity, and the chemical enrichment
history are strongly correlated with the stellar mass (e.g.,
\citealt{kauffmann2003}; \citealt{tremonti2004};
\citealt{gallazzi2005}). Furthermore, the evolution of the stellar
mass function (e.g., \citealt{bundy2006}; \citealt{marchesini2009};
\citealt{muzzin2013b}; \citealt{ilbert2013}) provides strong
constraints on galaxy formation models (see e.g.,
\citealt{delucia2007}).

In contrast to the luminosity, the stellar mass of a galaxy is not a
direct observable quantity.  Most techniques for estimating the
stellar mass rely on a determination of a mass-to-light ratio.  The
$M/L$ of a galaxy strongly depends on the age, metallicity, and the
stellar initial mass function (IMF) of its stellar
population. $M/L$ are typically estimated by comparing the observed colors, 
multi-wavelength broadband photometry or spectra to stellar population synthesis
(SPS) models (for a review see \citealt{conroy2013}).

In this Paper, we focus on the relation between the $M/L$ and color,
as was first explored by \citet{bell2001}. They used SPS models and
derived a tight relation between rest-frame $B-R$ color and $M/L_{\rm
B}$, from which it is possible to estimate the $M/L$ of a galaxy to an
accuracy of $\sim 0.2$ dex. Because their results were based on SPS
models, they suffer from uncertainties due to assumptions regarding
the star formation history (SFH), metallicity, IMF, and SPS code. More
recent work indeed suggests that the uncertainties are larger (0.2-0.4
dex; \citealt{bell2003}; \citealt{zibetti2009}; \citealt{taylor2011}),
in particular when using rest-frame NIR colors.

%

%
\begin{deluxetable*}{l l l l l l l }[!ht]

\tabletypesize{\scriptsize}
\tablecaption{Data References Sample}

\tablehead{\colhead{Survey \& Field} & \colhead{N$_{\rm gal}$} & \colhead{z} & \colhead{Spectroscopy} &
\colhead{Telescope \& } &  \colhead{Photometric Catalog}  & \colhead{Structural Parameters} \\
& & & & \colhead{Instrument} & & \\}

\vspace{0.2cm} 
SDSS DR7 & 4621 & $0.05<z< 0.07$ &\citet{abazajian2009} & SDSS & \citet{blanton2005}  & \citet{simard2011}  \\

NMBS-COSMOS & 3 & $0.7<z<0.9$ & \citet{bezanson2013b} & Keck-DEIMOS &\citet{skelton2014} & \citet{bezanson2011} \\
 & 10 & & & & \citet{whitaker2011} & \\

UKIDSS-UDS & 3 & $0.6<z<0.7$ & \citet{bezanson2013b} & Keck-DEIMOS& \citet{skelton2014} & \citet{vanderwel2012} \\
\vspace{0.2cm} 
& 1 & & & &\citet{williams2009} & \\

MS 1054-0321 & 8 & $z=0.83$ & \citet{wuyts2004}& Keck-LRIS & \citeauthor{forsterschreiber2006} & \citet{blakeslee2006} \\
& & & & & \citeyear{forsterschreiber2006} &  \\
\vspace{0.2cm}
GOODS-S & 7 & $ 0.9 < z < 1.2 $ & \citet{vanderwel2005}& VLT-FORS2 & \citet{skelton2014}  & \citet{vanderwel2012} \\

\vspace{0.2cm}
GOODS-N & 1 & $ z =1.315  $ & \citet{newman2010}& Keck-LRIS & \citet{skelton2014}  & \citet{vanderwel2012} \\

EGS & 8 & $1.0 < z< 1.3$ & \citet{belli2014a}& Keck-LRIS & \citet{skelton2014} & \citet{vanderwel2012} \\
COSMOS & 6 & $ 1.1 < z < 1.3 $ & \citet{belli2014a}& Keck-LRIS & \citet{skelton2014}  & \citet{vanderwel2012} \\
\vspace{0.2cm}
GOODS-S & 1 & $z = 1.419 $ & \citet{belli2014a} & Keck-LRIS & \citet{skelton2014} & \citet{vanderwel2012} \\

NMBS-COSMOS & 4 &  $1.2<z<1.5$ & \citet{bezanson2013a}& Keck-LRIS & \citet{whitaker2011}  & \citet{bezanson2013a}\\
\vspace{0.2cm}
NMBS-AEGIS & 2  & $1.4<z<1.6$ &  \citet{bezanson2013a} & Keck-LRIS& \citet{whitaker2011}  & \citet{bezanson2013a} \\

NMBS--COSMOS & 2 &  $1.6<z<2.1$ & \citet{vandesande2013} & VLT-XShooter &  \citet{skelton2014}  & \citet{vandesande2013} \\
 & 1 & & & & \citet{whitaker2011} & \\
UKIDSS-UDS & 1 &  $1.4<z<2.1$ & \citet{vandesande2013} & VLT-XShooter &  \citet{skelton2014}  & \citet{vandesande2013} \\
\vspace{0.2cm}
 &1 & & & &\citet{williams2009} & \\

COSMOS & 1 & $z=1.823 $ & \citet{onodera2012} & Subaru-MOIRCS & \citet{muzzin2013a} & \citet{onodera2012} \\
\vspace{0.2cm}
MUSYC 1255 & 1 &  $z=2.286$ &  \citet{vandokkum2009}&Gemini-GNIRS & \citet{blanc2008} &  \citet{vandokkum2009} \\

COSMOS & 2 & $ 2.1 < z < 2.3 $ 
& \citet{belli2014b}& Keck-MOSFIRE & \citet{skelton2014} &
\citet{belli2014b} \\ 

\enddata
\label{tab:tab1}
\end{deluxetable*}

%
\begin{deluxetable*}{c c c c c c c c c c c c c }[!ht]
\tabletypesize{\scriptsize}
\tablecaption{Absolute Magnitudes of the Sun in different Filters}
\tablehead{
 \colhead{$M_{\rm \odot,U}$}&\colhead{$M_{\rm \odot, B}$} &\colhead{$M_{\rm \odot,V}$} & \colhead{$M_{\rm \odot,R}$} & \colhead{$M_{\rm \odot,I}$} &
\colhead{$M_{\rm \odot,u}$} & \colhead{$M_{\rm \odot,g}$} & \colhead{$M_{\rm \odot,r}$} & \colhead{$M_{\rm \odot,i}$} &
\colhead{$M_{\rm \odot,z}$} & \colhead{$M_{\rm \odot,J}$}  & \colhead{$M_{\rm \odot,H}$} & \colhead{$M_{\rm \odot,K}$} \\}
6.34 & 5.33 & 4.81  & 4.65 & 4.55 & 6.45 &5.14 & 4.65 & 4.54 & 4.52 & 4.57 & 4.71& 5.19 \\
\enddata
\label{tab:tab2}
\end{deluxetable*}


Direct stellar kinematic mass measurements yield dynamical $M/L$,
which do not rely on any assumptions regarding the SPS models,
metallicity, and IMF. At low-redshift, dynamical mass measurements of
galaxies have proven to be extremely useful for studying the $M/L$
(e.g, \citealt{cappellari2006}; \citealt{dejong2007};
\citealt{taylor2010}). For example, \citet{cappellari2006} find that
the stellar $M/L$ is tightly correlated with $\sigma_e$, and
\citet{taylor2010} find that the stellar $M/L$ is a good predictor of
the dynamical $M/L$ if the the S\'ersic index is taken into account
when calculating dynamical masses (i.e., $M_{\rm dyn} \propto K(n)
r_{\rm e} \sigma_{\rm e}^2$).  \citet{vanderwel2006} studied the
relation between the dynamical $M/L_{\rm K}$ and rest-frame $B-K$
color of early-type galaxies out to $z\sim1$, and found that there are
large discrepancies between different SPS models in the NIR. However,
one of the major limitations in this measurement was the low number of
galaxies, and the small dynamic range in $M/L_{\rm K}$ ($\sim$0.4
dex).

In order to accurately constrain the relation between the dynamical
$M/L$ and color, we need a sample of early-type galaxies with a large
range in age. This study requires kinematic measurements from $z\sim0$
to $z\sim2$, such that we measure the $M/L$ and color in early-type
galaxies with both the oldest ($z\sim0$) and youngest ($z\sim2$)
stellar populations.

However, due to observational challenges very few such measurements
exist. At high-redshift, kinematic studies of quiescent galaxies
become much more difficult as the bulk of the stellar light, and
stellar absorption features used to measure velocity dispersions,
shift into the near-infrared (NIR) (e.g., \citealt{kriek2009}). With
the advent of fully depleted, high-resistivity CCDs (e.g., Keck-LRIS),
and new NIR spectrographs, such as VLT-X-SHOOTER \citep{vernet2011},
and Keck-MOSFIRE \citep{mclean2012}, it is now possible to obtain
rest-frame optical spectra of quiescent galaxies out to $z\sim2$.  For
example, \citet{bezanson2013a} measured accurate dynamical masses for
8 galaxies at $1.2<z<1.6$. Furthermore, in \citeauthor{vandesande2011} 
(\citeyear{vandesande2011}; \citeyear{vandesande2013}) we
obtained stellar kinematic measurements for 5 massive quiescent
galaxies up to redshift $z=2.1$ (see also \citealt{toft2012};
\citealt{belli2014b}). Combined with high-resolution imaging and
multi-wavelength catalogs, these recently acquired kinematic
measurements increase the dynamic range of the $M/L$ and rest-frame
color.

In this paper, we use a sample of massive quiescent galaxies from
$z\sim2$ to $z\sim0$ with kinematic measurements and multi-band
photometry with the aim of exploring the relation between the $M/L$
and rest-frame color, assessing SPS models, and constraining the IMF.
The paper is organized as follows. In Section \ref{sec:data} we
present our sample of $0.05 < z < 2.1$ galaxies, discuss the
photometric and spectroscopic data, and describe the derived galaxy
properties such as effective radii, rest-frame fluxes, and stellar
population parameters. In Section \ref{sec:emp} we explore the
relation between the $M/L$ and the rest-frame color over a large
dynamic range for several pass-bands.  We compare our $M/L$
vs. rest-frame color to predictions from stellar population models in
Section \ref{sec:comp}. In Section \ref{sec:imf} we use the
$M/L$ vs. color and stellar population models to constrain the IMF in
massive galaxies, as first proposed by \citeauthor{tinsley1972}
(\citeyear{tinsley1972}; \citeyear{tinsley1980}; see also
\citealt{vandokkum2008a} who first applied this technique to measure
the IMF out to $z\sim1$). In Section \ref{sec:disc} we compare
our results with previous measurements and discus several
uncertainties. Finally, in Section \ref{sec:conclusions} we
summarize our results and conclusions. Throughout the paper we assume
a $\Lambda$CDM cosmology with $\Omega_\mathrm{m}$=0.3
$\Omega_{\Lambda}=0.7$, and $H_{0}=70$ km s$^{-1}$ Mpc$^{-1}$. All
broadband data are given in the AB-based photometric system.


\begin{figure*}
\epsscale{0.9}
\plotone{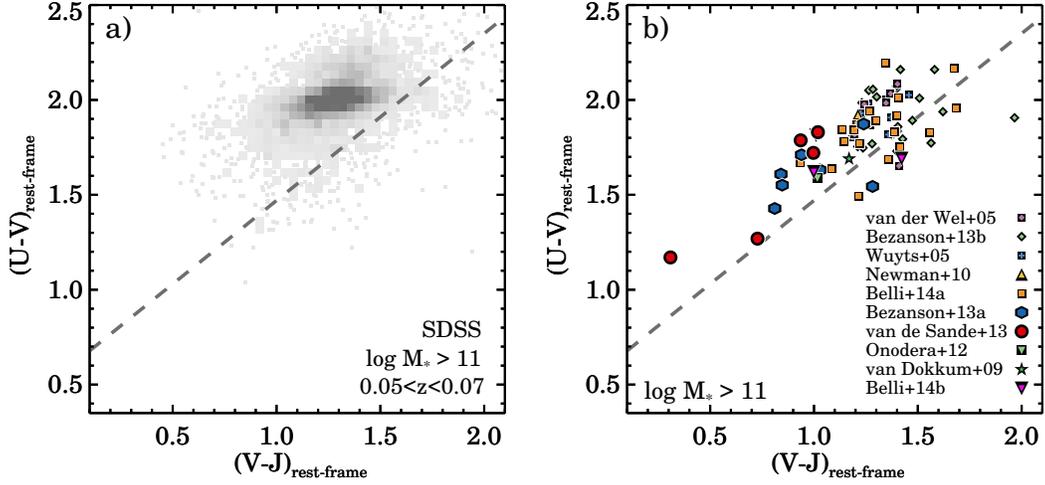}
\caption{Rest-frame $U-V$ color vs. $V-J$ color. Panel a) shows
massive ($M_* > 10^{11}$ \msun) galaxies in the SDSS at $z\sim0.06$,
and Panel b) shows massive galaxies at $z>0.5$. Different symbols for
the intermediate to high-redshift samples are indicated in the legend
and described in Section \ref{sec:data}. The dashed lines shows our
separation of star-forming and quiescent galaxies, where quiescent
galaxies are selected to have $U-V > (V-J ) \times 0.88 + 0.59$. We
only use the quiescent galaxies in the remainder of the Paper.
}
\label{fig:fig1}
\end{figure*}
\begin{figure*}
\epsscale{0.9}
\plotone{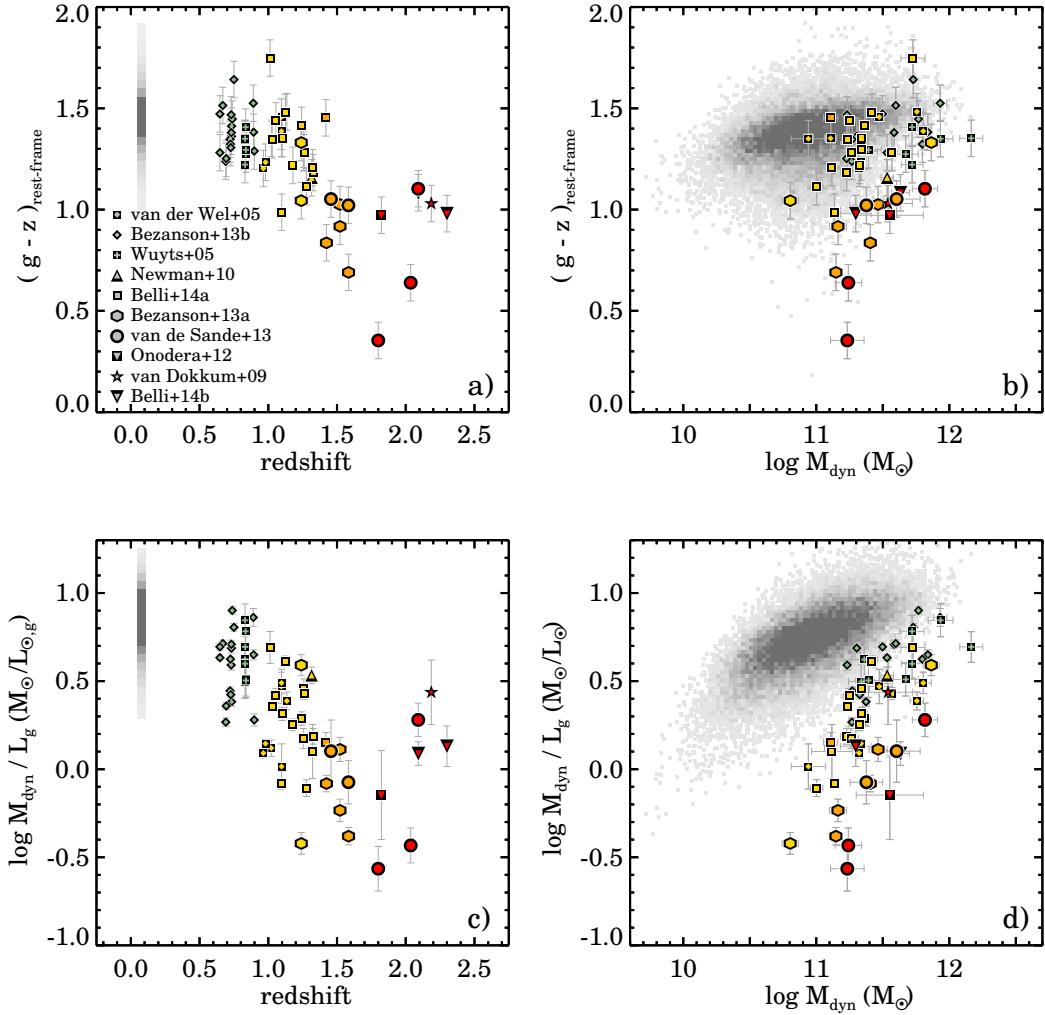}
\caption{Panel (a): Rest-frame $g-z$ color vs. redshift. Massive
quiescent galaxies in the SDSS at $z\sim0.06$ are represented by gray squares. 
Different symbols are the
different intermediate to high-redshift samples as indicated in the
legend and described in Section \ref{sec:data}. Galaxies are color
coded by redshift, from low (blue) to high redshift (red). We find
that the complete sample has a large range in colors. Panel (b): Rest-frame $g-z$ color versus
dynamical mass. Panel (c): $M_{\rm{dyn}}/L_{\rm g}$ in the
rest-frame g band vs. redshift. We find that the complete sample has a
large dynamic range with a factor of $\sim$25 in $M_{\rm{dyn}}/L_{\rm
g}$. Panel (d): $M_{\rm{dyn}}/L_{\rm g}$ versus dynamical mass.
Similar to panel b, we color code by redshift. 
From Panel (a) to (d) we conclude that, within the sample at fixed dynamical mass, 
the highest-redshift galaxies have the bluest colors with the lowest $M/L$.
}
\label{fig:fig2}
\end{figure*}
%


\section{Data}
\label{sec:data}

\subsection{Low and High-Redshift Sample}

One of the primary goals of this paper is to explore the relation
between the $M/L$ and rest-frame color over a large dynamic range.
Early-type galaxies are ideal candidates for such a measurement, as
they have homogeneous stellar populations. At $z\sim0$ their spectral
energy distributions are dominated by old stellar populations (e.g.,
\citealt{kuntschner2010}), and they have experienced very little to no
star formation since $z\sim2$ (e.g., \citealt{kriek2008}).

Here, we use a variety of datasets, which all contain stellar
kinematic measurements of individual galaxies and multi-wavelength
medium and broad-band photometric catalogs. We adopt a mass limit of
$M_* > 10^{11} $\msun to homogenize the final sample. Our
mass-selected sample contains 76 massive galaxies at $0.5<z<2.2$. We
note, however, that our sample remains relatively heterogeneous
relative to mass-complete photometric samples and in particular the
higher redshift samples are biased toward the brightest galaxies.

We use the $U-V$ versus $V-J$ rest-frame color selection to
distinguish quiescent galaxies from (dusty) star-forming galaxies.
(e.g., \citealt{wuyts2007}; \citealt{williams2009}). Figure
\ref{fig:fig1} shows the mass selected sample in the UVJ diagram, in
which quiescent galaxies have $U-V > (V-J ) \times 0.88 + 0.59$.  Out
of the 76 galaxies in the mass selected sample, 13 galaxies are not identified 
as quiescent galaxies and excluded from our sample. This criterion is
slightly different from previous work, as we do not require that $U-V
> 1.3$ or $V-J < 1.5$. The latter criteria remove
post-starburst galaxies and very old galaxies, respectively. As we benefit from a large
range in age in this Paper, we omit the latter criteria and thereby
keep the youngest and oldest galaxies in our sample.

Photometry for the high-redshift sample is adopted from the 3D-HST
catalogs version 4.1 (\citealt{brammer2012}; \citealt{skelton2014})
where possible, which cover the following CANDELS fields: AEGIS,
COSMOS, GOODS-N, GOODS-S and UKIDSS-UDS. We list the references for
all kinematic studies, photometric catalogs, and structural parameters
for the final sample in Table \ref{tab:tab1}.

\subsection{Derived Galaxy Properties}
\label{subsec:galprop}

All velocity dispersions were measured from stellar absorption
features in the rest-frame near-UV and/or optical.  We apply an
aperture correction to the velocity dispersion measurements as if they
were observed within a circular aperture radius of one $r_e$,
following the method as described in \citet{vandesande2013}. This
method includes a correction for the radial dependence of the velocity
dispersion (e.g., \citealt{cappellari2006}), and takes into account
the effects of the non-circular aperture, seeing, and optimal
extraction of the 1-D spectrum.

For the intermediate to high-redshift sample, effective radii and
other structural parameters, such as \ser index and axis ratio, are
determined using 2D S\'ersic fits with GALFIT \citep{peng2010}. For
galaxies in the SDSS, we use the structural parameters from
\citet{simard2011}, who determined 2D S\'ersic fits with GIMD2D
\citep{simard1998} on the SDSS $g$ band imaging data. All effective
radii are circularized, i.e., $r_{\rm{e}} = \sqrt{ab}$. All sizes are
measured from rest-frame optical data, i.e., redwards of 4000\AA, with
the exception of COSMOS-13412 ($z=1.24$) from \citet{bezanson2013a},
and COSMOS-254025 ($z=1.82$) from \citet{onodera2012} for which the
HST-F775W band is used. For massive galaxies at $z>1$, the median
color gradient is $r_{\rm e,u} / r_{\rm e,g} = 1.12$
\citep{szomoru2013}.  Thus the $M/L$ for these two galaxies may be
overestimated by $\sim0.05$ dex.

All rest-frame fluxes, including those for the SDSS sample, are
calculated using the photometric redshift code EAZY (v46;
\citealt{brammer2008}). We use the same set of templates that were
used for the ULTRAVISTA catalog by \citet{muzzin2013a}. Stellar masses
for the high-redshift sample are derived using the stellar population
fitting code FAST \citep{kriek2009}. We use the \citet{bruzual2003}
SPS models and assume an exponentially declining star formation
history (SFH), solar metallicity ($Z=0.02$), the \citet{calzetti2000}
dust attenuation law, and the \citet{chabrier2003} IMF.  For galaxies
in the SDSS, stellar masses are from the MPA-JHU
DR7\footnote{http://www.mpa-garching.mpg.de/SDSS/DR7/} release which
are based on \citet{brinchmann2004}, assuming a \citet{chabrier2003}
IMF.  The photometry and thus also the stellar mass are corrected for
missing flux using the best-fit S\'ersic luminosity
\citep{taylor2010}.

Dynamical masses are estimated from the size and velocity dispersion
measurements using the following expression:
\begin{equation} 
M_{\rm dyn}=\frac{\beta(n)~ \sigma_{\rm e}^2~r_{\rm e} }{G}.
\label{eq:mdyn}
\end{equation}
Here $\beta(n)$ is an analytic expression as a function of the
S\'ersic index, as described by \citet{cappellari2006}:
\begin{equation} 
\beta(n) = 8.87 - 0.831n + 0.0241n^2.
\label{eq:kn}
\end{equation}
We note that if we use a fixed virial constant of $\beta = 5$ for all
galaxies our conclusion would not change.

We derive mass-to-light ratios ($M/L_{\rm{\lambda}}$) using the
dynamical mass from Equation \ref{eq:mdyn} divided by the total
luminosity, in units of $M_{\odot} L_{\odot,\rm{\lambda}}^{-1}$. The
total luminosities for different wave bands ($\lambda$) are calculated
from rest-frame fluxes, as derived using EAZY. We normalize the total
luminosity using the absolute magnitude of the Sun in that particular
filter, which is measured from the solar spectrum taken from the
CALSPEC database
\footnote{http://www.stsci.edu/hst/observatory/cdbs/calspec.html}. The
solar absolute magnitudes for all filters are listed in Table
\ref{tab:tab2}.

%
\begin{figure*}
\epsscale{1.0}
\plotone{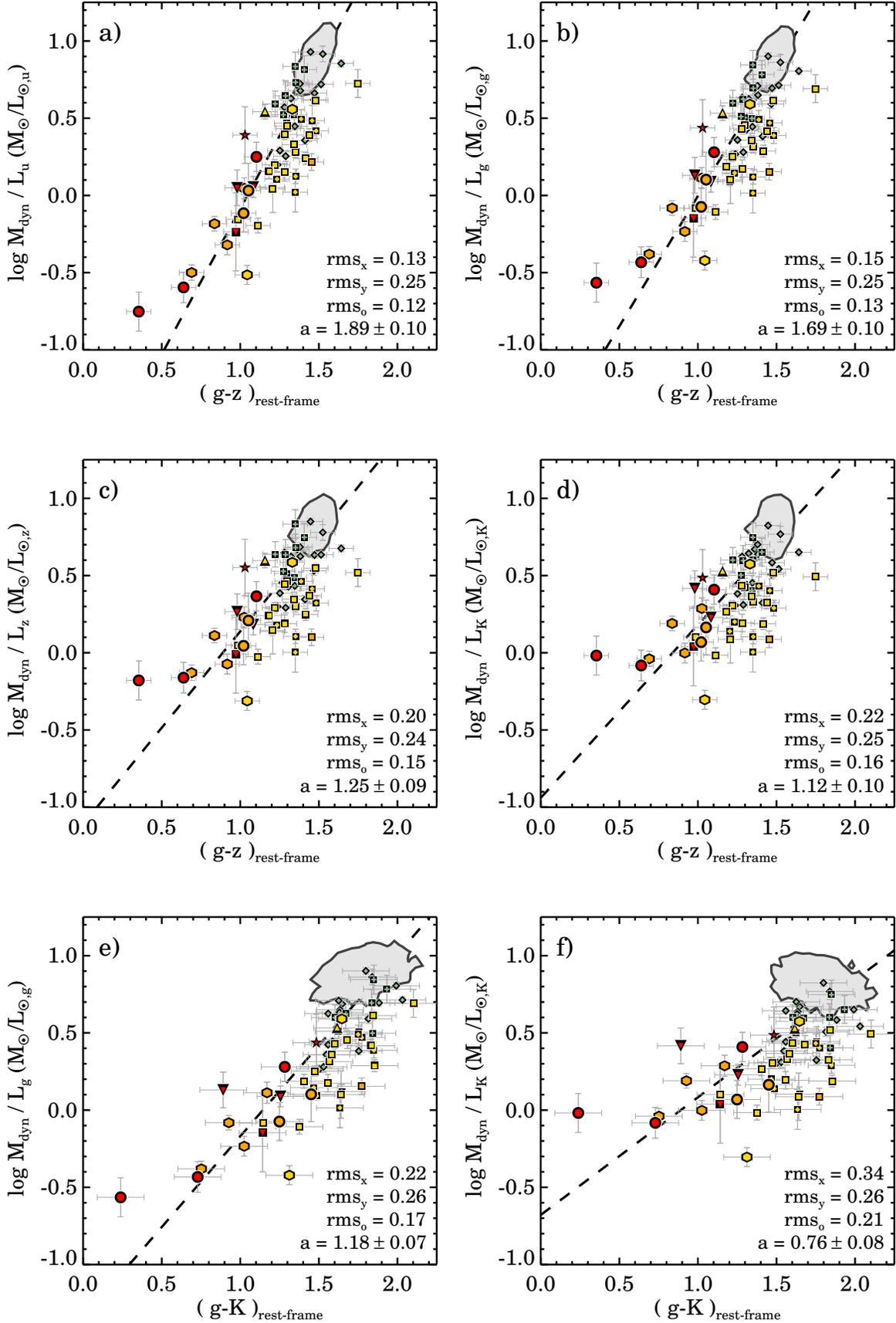}
\caption{$M/L$ vs. rest-frame color for massive galaxies between $z=0$
and $z=2$ for different luminosity bands and rest-frame
colors. Symbols are similar as in Figure \ref{fig:fig1}, and color
coded by redshift. From the rest-frame $u$ to the $K$ band, there is a
large range in the $M/L$: 1.8 dex in rest-frame $M/L_{\rm u}$, 1.6 dex
in $M/L_{\rm g}$, and 1.3 dex in $M/L_{\rm K}$. We find a strong
correlation between the $M/L$ for different luminosity bands and the
rest-frame $g-z$ color. The $M_{\rm{dyn}}/L_{\rm u}$ versus
$(g-z)_{\rm rest-frame}$ show the least amount of scatter. We use a
linear fit to the data to describe the relation between the $M/L$ and
rest-frame color, in which galaxies from the SDSS (gray
contour) are given equal weight as the high-redshift data. The
best-fit values are summarized in Table \ref{tab:tab3} \&
\ref{tab:tab4}. For each fit we furthermore give the rms scatter
orthogonal to the best-fitting line. Overall, the $M_{\rm{dyn}}/L_{\rm
g}$ vs $g-z$ color gives little scatter (rms$_o \sim 0.12$ dex),
whereas the scatter is higher when we use the rest-frame $g-K$ color
(rms$_o \sim 0.17$ dex).}
\label{fig:fig3}
\end{figure*}

\begin{figure}
\epsscale{1.1}
\plotone{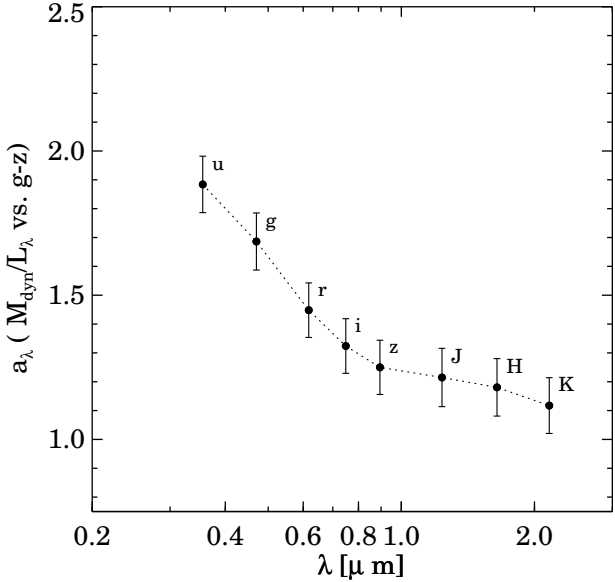}
\caption{Slope $a_{\rm \lambda}$ from the $M/L_{\rm \lambda}$ vs. \gz
relation versus wavelength $\lambda$. For each luminosity band, we use
the values from Table \ref{tab:tab3}, and the errors are derived from
bootstrapping the data. In the rest-frame $u$ band, we find a steep
slope of $\sim2.0$. The slope $a_{\rm \lambda}$ decreases when we go
from the $u$ to $i$ band. Redwards of the $z$ band, we find that the
relation between the $M/L$ vs. $g-z$ color is approximately
$\sim1.2$.}
\label{fig:fig4}
\end{figure}


\begin{table*}[h]
\centering
\rotatebox{90}{
\begin{minipage}{1.1\textwidth}
\caption{Emperical Relations for the $M/L$ vs. rest-frame Color using SDSS filters}
\setlength{\tabcolsep}{4pt}
\label{tab:data1}
\begin{tabular}{lccccccccccccccccccccccccccccccc}
\hline
\hline
\colhead{color} 
& \colhead{$a_{\rm u}$} & \colhead{$b_{\rm u}$} & \colhead{rms$_{\rm u}$} 
& \colhead{$a_{\rm g}$} & \colhead{$b_{\rm g}$} & \colhead{rms$_{\rm g}$} 
& \colhead{$a_{\rm r}$} & \colhead{$b_{\rm r}$} & \colhead{rms$_{\rm r}$} 
& \colhead{$a_{\rm i}$} & \colhead{$b_{\rm i}$} & \colhead{rms$_{\rm i}$} 
& \colhead{$a_{\rm z}$} & \colhead{$b_{\rm z}$} & \colhead{rms$_{\rm z}$} 
& \colhead{$a_{\rm J}$} & \colhead{$b_{\rm J}$} & \colhead{rms$_{\rm J}$}
& \colhead{$a_{\rm H}$} & \colhead{$b_{\rm H}$} & \colhead{rms$_{\rm H}$}
& \colhead{$a_{\rm K}$} & \colhead{$b_{\rm K}$} & \colhead{rms$_{\rm K}$} \\
\hline
\\
\hline
    $       u-g       $ &    5.02    &   -7.35    &    0.08    &    4.77    &   -6.95    &    0.08    &    4.31    &   -6.21    &    0.09    &    4.08    &   -5.85    &    0.09    &    4.10    &   -5.91    &    0.09    &    4.07    &   -5.89    &    0.09    &    4.19    &   -6.08    &    0.10    &    4.04    &   -5.81    &    0.10   \\
    $       u-r       $ &    1.93    &   -3.92    &    0.12    &    1.72    &   -3.42    &    0.13    &    1.48    &   -2.86    &    0.14    &    1.36    &   -2.58    &    0.15    &    1.29    &   -2.45    &    0.15    &    1.26    &   -2.38    &    0.16    &    1.22    &   -2.29    &    0.16    &    1.16    &   -2.14    &    0.16   \\
    $       u-i       $ &    1.47    &   -3.31    &    0.33    &    1.31    &   -2.87    &    0.31    &    1.13    &   -2.39    &    0.30    &    1.04    &   -2.15    &    0.30    &    0.98    &   -2.03    &    0.29    &    0.96    &   -1.99    &    0.28    &    0.93    &   -1.92    &    0.28    &    0.89    &   -1.78    &    0.28   \\
    $       u-z       $ &    1.28    &   -3.18    &    0.66    &    1.14    &   -2.76    &    0.61    &    0.99    &   -2.29    &    0.58    &    0.91    &   -2.05    &    0.56    &    0.85    &   -1.93    &    0.55    &    0.83    &   -1.87    &    0.53    &    0.81    &   -1.82    &    0.53    &    0.77    &   -1.68    &    0.53   \\
    $       u-J       $ &    1.09    &   -3.00    &    1.08    &    0.97    &   -2.58    &    1.02    &    0.83    &   -2.11    &    0.97    &    0.76    &   -1.88    &    0.94    &    0.71    &   -1.75    &    0.92    &    0.67    &   -1.65    &    0.90    &    0.65    &   -1.60    &    0.90    &    0.62    &   -1.47    &    0.89   \\
    $       u-H       $ &    1.04    &   -3.04    &    1.30    &    0.93    &   -2.62    &    1.23    &    0.80    &   -2.17    &    1.16    &    0.73    &   -1.93    &    1.13    &    0.69    &   -1.81    &    1.11    &    0.65    &   -1.70    &    1.09    &    0.63    &   -1.64    &    1.08    &    0.60    &   -1.51    &    1.07   \\
    $       u-K       $ &    0.99    &   -2.61    &    0.96    &    0.88    &   -2.26    &    0.92    &    0.76    &   -1.85    &    0.88    &    0.70    &   -1.65    &    0.86    &    0.66    &   -1.55    &    0.84    &    0.63    &   -1.48    &    0.83    &    0.61    &   -1.42    &    0.83    &    0.57    &   -1.28    &    0.82   \\
    $       g-r       $ &    4.06    &   -2.47    &    0.07    &    3.73    &   -2.22    &    0.08    &    3.46    &   -2.01    &    0.09    &    3.24    &   -1.83    &    0.09    &    3.24    &   -1.86    &    0.09    &    3.30    &   -1.93    &    0.10    &    3.22    &   -1.87    &    0.10    &    3.14    &   -1.80    &    0.10   \\
    $       g-i       $ &    2.62    &   -2.22    &    0.10    &    2.41    &   -1.99    &    0.11    &    1.92    &   -1.46    &    0.12    &    1.75    &   -1.27    &    0.13    &    1.65    &   -1.19    &    0.13    &    1.61    &   -1.16    &    0.14    &    1.56    &   -1.12    &    0.14    &    1.47    &   -1.00    &    0.14   \\
    $       g-z       $ &    1.89    &   -1.97    &    0.12    &    1.69    &   -1.69    &    0.13    &    1.45    &   -1.36    &    0.14    &    1.32    &   -1.18    &    0.15    &    1.25    &   -1.11    &    0.15    &    1.21    &   -1.08    &    0.16    &    1.18    &   -1.04    &    0.16    &    1.12    &   -0.94    &    0.16   \\
    $       g-J       $ &    1.54    &   -2.03    &    0.16    &    1.37    &   -1.74    &    0.16    &    1.18    &   -1.40    &    0.17    &    1.07    &   -1.21    &    0.18    &    1.00    &   -1.12    &    0.19    &    0.95    &   -1.05    &    0.20    &    0.92    &   -1.01    &    0.20    &    0.87    &   -0.91    &    0.20   \\
    $       g-H       $ &    1.46    &   -2.18    &    0.16    &    1.31    &   -1.88    &    0.17    &    1.13    &   -1.52    &    0.18    &    1.03    &   -1.34    &    0.18    &    0.97    &   -1.24    &    0.19    &    0.92    &   -1.17    &    0.20    &    0.89    &   -1.12    &    0.20    &    0.84    &   -1.01    &    0.20   \\
    $       g-K       $ &    1.31    &   -1.57    &    0.17    &    1.18    &   -1.35    &    0.17    &    1.02    &   -1.07    &    0.18    &    0.93    &   -0.93    &    0.18    &    0.88    &   -0.87    &    0.19    &    0.84    &   -0.83    &    0.20    &    0.82    &   -0.79    &    0.20    &    0.76    &   -0.68    &    0.21   \\
    $       r-i       $ &    7.33    &   -1.71    &    0.04    &    7.05    &   -1.63    &    0.04    &    6.66    &   -1.50    &    0.05    &    6.42    &   -1.42    &    0.05    &    6.41    &   -1.44    &    0.05    &    5.91    &   -1.29    &    0.05    &    6.18    &   -1.40    &    0.05    &    6.11    &   -1.36    &    0.05   \\
    $       r-z       $ &    4.87    &   -2.34    &    0.06    &    4.52    &   -2.11    &    0.07    &    4.30    &   -1.96    &    0.07    &    4.00    &   -1.80    &    0.08    &    3.95    &   -1.80    &    0.08    &    4.13    &   -1.91    &    0.09    &    4.12    &   -1.91    &    0.09    &    3.87    &   -1.74    &    0.09   \\
    $       r-J       $ &    2.57    &   -1.85    &    0.13    &    2.28    &   -1.56    &    0.13    &    1.93    &   -1.22    &    0.14    &    1.75    &   -1.05    &    0.15    &    1.60    &   -0.93    &    0.16    &    1.48    &   -0.83    &    0.17    &    1.43    &   -0.79    &    0.17    &    1.35    &   -0.69    &    0.18   \\
    $       r-H       $ &    2.34    &   -2.06    &    0.13    &    2.07    &   -1.74    &    0.14    &    1.75    &   -1.37    &    0.15    &    1.59    &   -1.19    &    0.15    &    1.48    &   -1.10    &    0.16    &    1.35    &   -0.97    &    0.17    &    1.30    &   -0.91    &    0.18    &    1.23    &   -0.81    &    0.18   \\
    $       r-K       $ &    1.99    &   -1.19    &    0.14    &    1.77    &   -0.99    &    0.15    &    1.53    &   -0.76    &    0.15    &    1.39    &   -0.64    &    0.16    &    1.29    &   -0.57    &    0.17    &    1.20    &   -0.51    &    0.18    &    1.16    &   -0.48    &    0.18    &    1.06    &   -0.37    &    0.19   \\
\hline
\label{tab:tab3}
\end{tabular}
%
\caption{Emperical Relations for the $M/L$ vs. rest-frame Color using Johnson-Cousin Filters}
\setlength{\tabcolsep}{4pt}
\label{tab:data2}
\begin{tabular}{lccccccccccccccccccccccccccccccc}
\hline
\hline
\colhead{color} 
& \colhead{$a_{\rm U}$} & \colhead{$b_{\rm U}$} & \colhead{rms$_{\rm U}$} 
& \colhead{$a_{\rm B}$} & \colhead{$b_{\rm B}$} & \colhead{rms$_{\rm B}$} 
& \colhead{$a_{\rm V}$} & \colhead{$b_{\rm V}$} & \colhead{rms$_{\rm V}$} 
& \colhead{$a_{\rm R}$} & \colhead{$b_{\rm R}$} & \colhead{rms$_{\rm R}$} 
& \colhead{$a_{\rm I}$} & \colhead{$b_{\rm I}$} & \colhead{rms$_{\rm I}$} 
& \colhead{$a_{\rm J}$} & \colhead{$b_{\rm J}$} & \colhead{rms$_{\rm J}$}
& \colhead{$a_{\rm H}$} & \colhead{$b_{\rm H}$} & \colhead{rms$_{\rm H}$}
& \colhead{$a_{\rm K}$} & \colhead{$b_{\rm K}$} & \colhead{rms$_{\rm K}$} \\
\hline
\\
\hline
     $       U-B       $ &   10.16    &  -10.42    &    0.06    &    9.86    &  -10.04    &    0.06    &    9.44    &   -9.62    &    0.06    &    9.32    &   -9.49    &    0.06    &    8.63    &   -8.73    &    0.06    &   10.31    &  -10.63    &    0.07    &    9.39    &   -9.59    &    0.07    &    9.19    &   -9.38    &    0.07   \\
    $       U-V       $ &    2.44    &   -4.12    &    0.10    &    2.28    &   -3.77    &    0.11    &    1.97    &   -3.20    &    0.12    &    1.83    &   -2.94    &    0.13    &    1.68    &   -2.65    &    0.13    &    1.58    &   -2.50    &    0.14    &    1.55    &   -2.44    &    0.14    &    1.47    &   -2.27    &    0.14   \\
    $       U-R       $ &    1.77    &   -3.47    &    0.13    &    1.66    &   -3.18    &    0.13    &    1.44    &   -2.69    &    0.14    &    1.33    &   -2.46    &    0.15    &    1.22    &   -2.22    &    0.16    &    1.15    &   -2.08    &    0.17    &    1.12    &   -2.01    &    0.17    &    1.06    &   -1.86    &    0.17   \\
    $       U-I       $ &    1.39    &   -3.05    &    0.30    &    1.31    &   -2.79    &    0.29    &    1.13    &   -2.37    &    0.28    &    1.05    &   -2.15    &    0.28    &    0.96    &   -1.93    &    0.28    &    0.90    &   -1.81    &    0.27    &    0.88    &   -1.75    &    0.27    &    0.84    &   -1.62    &    0.27   \\
    $       U-J       $ &    1.08    &   -2.84    &    0.94    &    1.01    &   -2.58    &    0.91    &    0.87    &   -2.17    &    0.86    &    0.80    &   -1.95    &    0.84    &    0.73    &   -1.73    &    0.82    &    0.67    &   -1.56    &    0.79    &    0.65    &   -1.50    &    0.79    &    0.61    &   -1.38    &    0.78   \\
    $       U-H       $ &    1.04    &   -2.90    &    1.15    &    0.97    &   -2.63    &    1.12    &    0.84    &   -2.22    &    1.06    &    0.78    &   -2.01    &    1.03    &    0.71    &   -1.78    &    1.01    &    0.65    &   -1.61    &    0.97    &    0.63    &   -1.55    &    0.97    &    0.59    &   -1.42    &    0.96   \\
    $       U-K       $ &    0.98    &   -2.47    &    0.83    &    0.92    &   -2.25    &    0.81    &    0.80    &   -1.90    &    0.78    &    0.74    &   -1.72    &    0.76    &    0.68    &   -1.52    &    0.75    &    0.62    &   -1.39    &    0.72    &    0.61    &   -1.33    &    0.72    &    0.57    &   -1.19    &    0.71   \\
    $       B-V       $ &    3.95    &   -2.75    &    0.07    &    3.79    &   -2.59    &    0.07    &    3.44    &   -2.30    &    0.08    &    3.27    &   -2.17    &    0.09    &    3.11    &   -2.05    &    0.09    &    3.22    &   -2.17    &    0.10    &    3.13    &   -2.10    &    0.10    &    2.94    &   -1.91    &    0.10   \\
    $       B-R       $ &    2.55    &   -2.52    &    0.10    &    2.47    &   -2.38    &    0.10    &    2.15    &   -2.01    &    0.11    &    1.83    &   -1.63    &    0.12    &    1.66    &   -1.44    &    0.13    &    1.58    &   -1.36    &    0.14    &    1.54    &   -1.32    &    0.14    &    1.47    &   -1.21    &    0.14   \\
    $       B-I       $ &    1.72    &   -2.06    &    0.12    &    1.63    &   -1.88    &    0.13    &    1.42    &   -1.57    &    0.14    &    1.30    &   -1.41    &    0.15    &    1.20    &   -1.25    &    0.15    &    1.14    &   -1.19    &    0.16    &    1.11    &   -1.15    &    0.16    &    1.05    &   -1.05    &    0.17   \\
    $       B-J       $ &    1.32    &   -2.16    &    0.17    &    1.24    &   -1.96    &    0.17    &    1.07    &   -1.64    &    0.18    &    0.99    &   -1.47    &    0.18    &    0.90    &   -1.29    &    0.19    &    0.82    &   -1.16    &    0.21    &    0.80    &   -1.12    &    0.21    &    0.76    &   -1.01    &    0.21   \\
    $       B-H       $ &    1.26    &   -2.27    &    0.17    &    1.19    &   -2.07    &    0.17    &    1.03    &   -1.75    &    0.18    &    0.95    &   -1.57    &    0.19    &    0.87    &   -1.38    &    0.19    &    0.80    &   -1.26    &    0.21    &    0.77    &   -1.20    &    0.21    &    0.73    &   -1.09    &    0.21   \\
    $       B-K       $ &    1.14    &   -1.73    &    0.17    &    1.08    &   -1.56    &    0.18    &    0.94    &   -1.31    &    0.18    &    0.87    &   -1.17    &    0.19    &    0.79    &   -1.02    &    0.19    &    0.74    &   -0.94    &    0.21    &    0.71    &   -0.90    &    0.21    &    0.67    &   -0.79    &    0.21   \\
    $       V-R       $ &    7.86    &   -2.31    &    0.04    &    8.00    &   -2.31    &    0.04    &    7.29    &   -2.08    &    0.04    &    6.96    &   -1.95    &    0.05    &    6.57    &   -1.84    &    0.05    &    6.87    &   -1.98    &    0.05    &    6.73    &   -1.91    &    0.05    &    6.35    &   -1.76    &    0.05   \\
    $       V-I       $ &    3.89    &   -2.09    &    0.07    &    3.75    &   -1.95    &    0.07    &    3.45    &   -1.78    &    0.08    &    3.23    &   -1.63    &    0.08    &    3.15    &   -1.59    &    0.09    &    3.23    &   -1.68    &    0.10    &    3.12    &   -1.59    &    0.10    &    3.02    &   -1.52    &    0.10   \\
    $       V-J       $ &    1.97    &   -1.85    &    0.14    &    1.85    &   -1.66    &    0.15    &    1.59    &   -1.37    &    0.15    &    1.45    &   -1.21    &    0.16    &    1.31    &   -1.04    &    0.17    &    1.15    &   -0.87    &    0.19    &    1.11    &   -0.83    &    0.19    &    1.05    &   -0.73    &    0.19   \\
    $       V-H       $ &    1.88    &   -2.09    &    0.14    &    1.76    &   -1.88    &    0.15    &    1.52    &   -1.56    &    0.16    &    1.39    &   -1.39    &    0.16    &    1.27    &   -1.22    &    0.17    &    1.12    &   -1.04    &    0.19    &    1.08    &   -0.98    &    0.19    &    1.02    &   -0.87    &    0.19   \\
    $       V-K       $ &    1.63    &   -1.35    &    0.16    &    1.54    &   -1.20    &    0.16    &    1.34    &   -0.99    &    0.16    &    1.23    &   -0.87    &    0.17    &    1.13    &   -0.75    &    0.17    &    1.02    &   -0.66    &    0.19    &    0.99    &   -0.62    &    0.19    &    0.91    &   -0.51    &    0.20   \\
\hline
\label{tab:tab4}
\end{tabular}
\end{minipage}
} 
\end{table*}

\section{Empirical Relation Between the $M/L$ and Color}
\label{sec:emp}
\subsection{Color and the $M/L$ Evolution}
\label{subsec:col_ml_evo}
In Figure \ref{fig:fig2}(a) and (c) we show the rest-frame $g-z$ color and
$M_{\rm dyn}/L_{\rm g}$ as a function of redshift. We find a large range
in $g-z$ color ($\sim1$mag) and $\log_{10} M_{\rm dyn}/L_{\rm g}$
($\sim1.6$ dex). At $z>1$, massive galaxies are bluer and have lower
$M/L$ as compared to low-redshift. In \citet{vandesande2014}
we showed that our high-redshift spectroscopic sample is biased toward young
quiescent galaxies. While this bias complicated the
analysis of the fundamental plane as presented in that work, here we
take advantage of that same bias, as it enables us to study massive
quiescent galaxies with a large range in stellar population
properties.

We show \gz versus the dynamical mass in Figure \ref{fig:fig2}(b). We
find a weak trend between dynamical mass and color for low-redshift
galaxies. At $z>1.5$ the lowest mass galaxies have the bluest
colors. Figure \ref{fig:fig2}(d) shows the \mlg versus the dynamical
mass. For galaxies in the SDSS, there is a positive correlation such
that low mass galaxies also have lower $M/L$ as compared to more
massive galaxies. In the mass range of $10^{11} < M_{\rm dyn}/
M_{\odot} < 10^{12}$, the \mlg for galaxies in the SDSS increases by 
about $\sim0.2~$dex. For galaxies at $z>0.5$ in our sample,
we find that galaxies with high $M/L$ are on average more massive as
compared to galaxies with low $M/L$.

\subsection{Empirical Relation Between the $M/L$ and Color}
\label{subsec:empfit}
Next, we examine empirical relations between the $M/L$ and color, as
first predicted by \citeauthor{tinsley1972} (\citeyear{tinsley1972};
\citeyear{tinsley1980}). In Figure \ref{fig:fig3}(a-d) we show the
dynamical $M/L$ vs. rest-frame $g-z$ color in the following filters:
$u$, $g$, $z$, and $K$. Figures \ref{fig:fig3}(e) and (f) show the rest-frame
$g-K$ colors versus the $M/L_{\rm g}$ and $M/L_{\rm K}$. Symbols are
similar to Figure \ref{fig:fig2}. Massive quiescent galaxies
($>10^{11} M_{\odot}$) from the SDSS are shown by the gray contour,
which encloses $68\%$ of all galaxies. We find that the $\log_{10}
M/L_{\rm u}$ varies most, from $\sim-0.7$ to $\sim1.1$. The range in
$\log_{10} M/L$ slowly decreases with increasing wavelength from $1.8$
dex in the $u$ band, to 1.6 dex in the $g$ band, and 1.3 dex in the
$z$ and $K$ band.

As expected, we find a strong positive correlation between the $M/L$
in all passbands and rest-frame colors. Following \citet{bell2001}, we
fit the simple relation:
\begin{equation} 
\log_{10} M_{\rm dyn}~/~L_{\rm \lambda} = a_{\rm \lambda}*{color}~+~b_{\rm \lambda}
\label{eq:ml}
\end{equation}

We use the IDL routine $LINMIX\_ERR$ which is a Bayesian method to measure the linear regression,
with regression coefficients $a_{\rm \lambda}$ and $b_{\rm \lambda}$. The advantage of using a Bayesian approach 
over a routine that minimizes $\chi^2$, is that the Bayesian approach incorporates the intrinsic scatter
as a fit parameter.
In the fit we give equal weight to the SDSS galaxy sample and our sample of galaxies at
intermediate to high redshift, instead of anchoring the fit to the
median of the SDSS galaxies. In practice this means that we add 63
galaxies to the intermediate and high-z sample, which have a $M/L$ and color equal to the
median of all SDSS galaxies. 
We note that when we use the IDL routine $FITEXY$, which minimizes the $\chi^2$ in the fit for both
$x$ (rest-frame color) and $y$ ($M/L$), we find on average $\sim5\%$ higher values for $a_{\rm \lambda}$.
The results are summarized in Table
\ref{tab:tab3} using the SDSS filters and in Table \ref{tab:tab4}
where we use the Johnson-Cousin Filters.

Besides the coefficients, we also report the root-mean-square (rms$_o$)
scatter, which is a good indicator for the significance of each
relation. The scatter around the linear relation increases from 0.12
to 0.16 when going from rest-frame optical $M/L_{\rm u}$ to rest-frame
near-infrared $M/L_{\rm K}$. Furthermore, we find that for the \mlg the
scatter is lower when we use the rest-frame $g-z$ color as compared to
the rest-frame $g-K$ color.

For the color \gz, we find that the slope of the relation becomes
flatter from the UV to the near-infrared. To investigate this trend in
more detail, we plot the slope $a_{\lambda}$ as a function of
wavelength from the $M/L_{\rm \lambda}$ vs. \gz relation in Figure
\ref{fig:fig4}. We find that the slope becomes flatter when we go from
the $u$ to $i$ waveband, while the slope is approximately constant
($\sim1.2$) from $z$ to $K$.

These results indicate that the $M/L$ of an early-type galaxy can be
predicted by a singe rest-frame color $g-z$ to an accuracy of
$\sim0.25$ dex.  In particular the rest-frame $g-z$ color in
combination with the rest-frame $g$ band luminosity, provides a good
constrain for mass measurements of quiescent galaxies, as it has a large 
dynamic range in $M/L$ and color, in combination
with little scatter around the linear relation.

\begin{figure*}
\epsscale{1.0}
\plotone{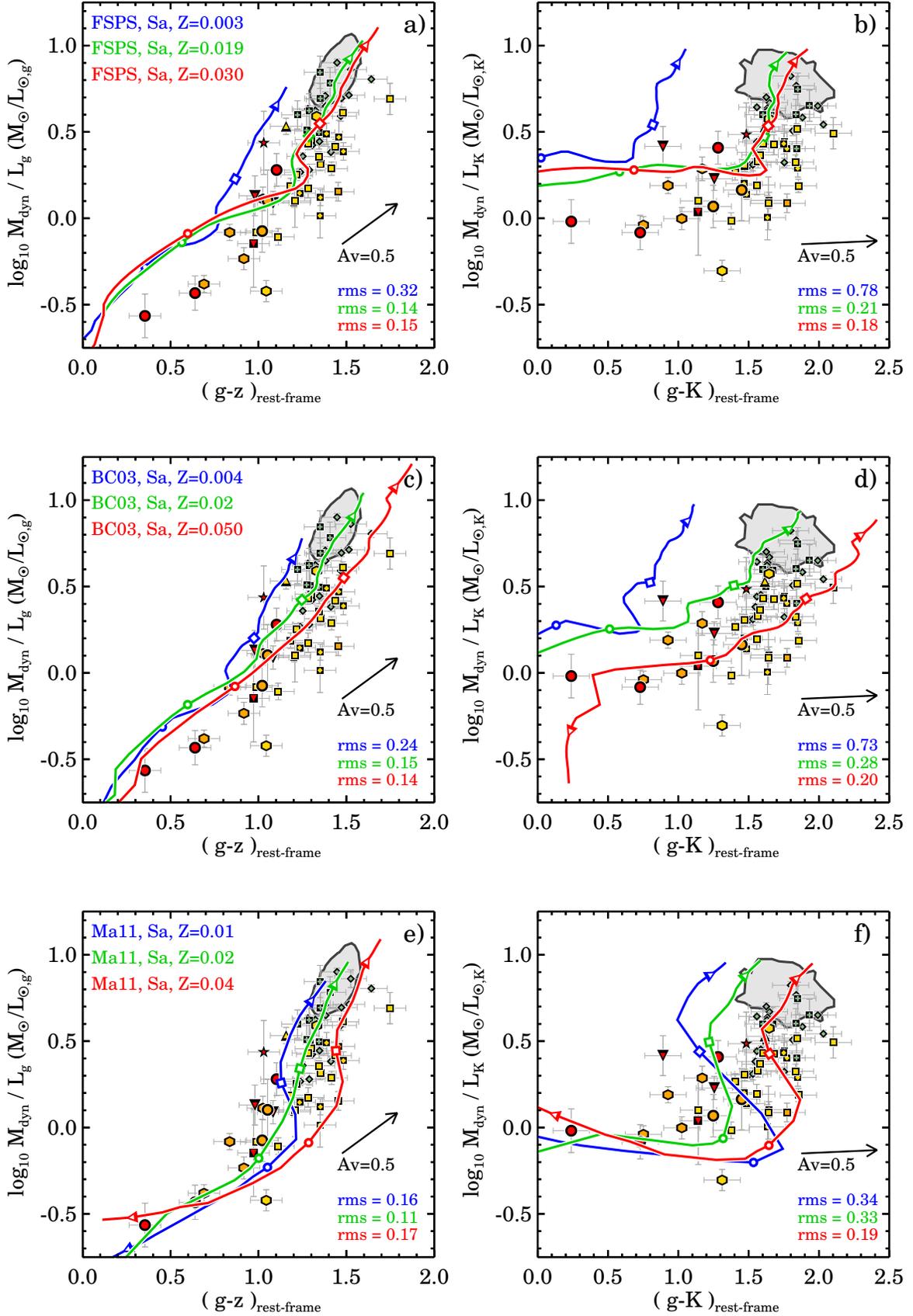}
\caption{$M/L$ vs. rest-frame color for the spectroscopic samples
compared to the following SPS models: FSPS (top row), BC03 (middle
row), and Ma11 (bottom row). The left column shows the \mlg vs. \gz,
while the right column shows the \mlk vs. \gk. For each model we show
three different metallicities: solar metallicity ($Z=0.02$, green),
sub-solar metallicity (blue) and super-solar (red), and we indicate
various model ages on the tracks with different symbols: 0.1 Gyr
(upside down triangle), 1.0 Gyr (circle), 3.0 Gyr (diamond), 10 Gyr
(triangle). The effect of dust is indicated by the black arrow. We
find that none of the models are able to match both the rest-frame
optical color and the rest-frame infrared color in combination with
the $M/L$ simultaneously.}
\label{fig:fig5}
\end{figure*}
%


\section{Comparisons with Stellar Population Synthesis Models}
\label{sec:comp}

Here, we compare our dynamical $M/L$ vs. rest-frame color to the
predictions from stellar population synthesis (SPS) models. Our main aim is
to test whether the different SPS models can reproduce the relations
in the optical and NIR and to what accuracy. For the comparison we
use the SPS models by \citeauthor{bruzual2003}
(\citeyear{bruzual2003}; BC03), \citeauthor{maraston2011}
(\citeyear{maraston2011}; Ma11), and \citet{conroy2010} (FSPS,
v2.4). 

For the BC03 models we use the simple stellar population (SSP) models with the
Padova stellar evolution tracks \citep{bertelli1994}, and the STELIB
stellar library \citep{leborgne2003}. For the Ma11 models, which are
based on \citeauthor{maraston2005} (\citeyear{maraston2005}; Ma05)
with the \citet{cassisi1997a}, \citet{cassisi1997b}, and
\citet{cassisi2000} stellar evolution tracks and isochrones, we use
SSPs with the red horizontal branch morphology and the MILES
\citep{sanchez2006} stellar library. For the FSPS models, which use
the latest Padova stellar evolution tracks (\citealt{marigo2007};
\citealt{marigo2008}), we use the standard program settings, and the
MILES stellar library.

For all models we use a \citet{salpeter1955} IMF and a truncated SFH
with a constant star formation rate for the first 0.5 Gyr. However,
different SFHs result in nearly identical tracks. For example a longer
star formation timescale will smooth out some of the small time-scale
variations, but will not change any of our conclusions. For all models
we use the total stellar mass, which is the sum of living stars and
remnants. We note that our dynamical mass estimates include both
stellar mass and dark matter mass. At this point we ignore the effect
of dark matter, but we come back to this issue in Section
\ref{subsec:dm}.

In Figure \ref{fig:fig5} we compare the \mlg vs. \gz (left column) and
\mlk vs. \gk (right column) with the predictions from SPS models. A different model is shown in each row
from top to bottom: FSPS, BC03, and Ma11. For each model, we show
three different metallicities: solar (green), sub-solar (blue) and
super-solar (red). Metallicity values for the specific models are
indicated in each panel. Furthermore, we indicate various model ages
on the tracks with different symbols: 0.1 Gyr (upside down triangle),
1.0 Gyr (circle), 3.0 Gyr (diamond), 10 Gyr (triangle).

We indicate the effect of dust with the black arrow, assuming a
\citet{calzetti2000} dust law and $A_{\rm V}=0.5$. For the \mlg
vs. rest-frame $g-z$ color, we find that the dust vector runs parallel
to the model. Dust has, however, almost no affect on the rest-frame
$K$ band luminosity and therefore runs nearly horizontal.  To quantify
how well the models match the data, we calculate the scatter
orthogonal to the model tracks (rms$_o$). For the three models with
different metallicities this scatter is given each panel in Figure \ref{fig:fig5} 
and in Table \ref{tab:tab5}. Below we discuss the comparison for SPS models 
individually.

\vspace{0.5cm}
\subsection{FSPS models}
\label{subsec:fsps}
In Figure \ref{fig:fig5}(a) we show the \mlg vs. \gz color in
combination with the FSPS models. The difference between the solar
($Z=0.02$) and super-solar ($Z=0.03$) tracks is small, whereas the
sub-solar model is at all times too blue at fixed $M/L$. Both models
with solar and super-solar metallicity match the \mlg and \gz color for
low-redshift galaxies ($z<1$), but are unable to reproduce the low
\mlg for the bluest galaxies with \gz$<1$. It is interesting to note
that the scatter around the Solar metallicity model (0.14, Table
\ref{tab:tab5}) is about similar to the scatter when assuming the
linear fit (0.13, Figure \ref{fig:fig3}(b)).

In Figure \ref{fig:fig5}(b) we show the \mlk vs. the \gk color. The FSPS
model tracks show a clear transition from a constant \mlk for \gk$<
1.75$, to a very steep relation at \gk$> 1.75$. The color difference
between the solar ($Z=0.02$) and sub-solar ($Z=0.003$) tracks is large
($\Delta $\gz$ \sim 0.8$) and more distinct as compared to the
rest-frame $g-z$ color. The difference between solar and super-solar
metallicity tracks is small for the FSPS models. The scatter for the Super-Solar metallicity
model (0.18) is similar to the scatter for the linear fit (0.21,
Figure \ref{fig:fig3}(f)), even though the FSPS model track is far from
linear. As in Figure \ref{fig:fig5}(a), we find that both solar and super-solar
tracks are able to match the low-redshift galaxies ($z<1$), but cannot
simultaneously match the bluest galaxies with \gk$ < 1.5$ colors, for
which the model $M/L$ is too high.


\begin{table}[!t]
\caption{Scatter around SPS Models with different metallicities}
\label{tab:data3}
\begin{center}
\begin{tabular}{l l l | c c c c}
\hline
\hline
\colhead{SPS Model} & \colhead{$M/L$} & \colhead{color}  & \colhead{Sub-Solar} & \colhead{Solar} & \colhead{Super-Solar} \\
\hline
FSPS     & $M/L_{\rm g}$   & $ g-z$ & 0.32 & 0.14 & 0.15 \\
            & $M/L_{\rm K}$   & $g-K$ & 0.78 & 0.21 & 0.18 \\
                          
BC03     & $M/L_{\rm g}$   & $g-z$ & 0.24 & 0.15 & 0.14 \\
         & $M/L_{\rm K}$   & $g-K$ & 0.73 & 0.28 & 0.20  \\
                          
Ma11     & $M/L_{\rm g}$   & $g-z$ & 0.16 & 0.11 & 0.17 \\
         & $M/L_{\rm K}$    & $g-K$ & 0.34 & 0.33 & 0.19 \\
\hline
\label{tab:tab5}
\end{tabular}
\end{center}
\end{table}

\subsection{BC03 models}
\label{subsec:bc03}
For the BC03 models, the solar-metallicity track matches the
low-redshift galaxies well for the rest-frame optical colors (Figure
\ref{fig:fig5}(c)), but the $M/L$ of the bluest galaxies is still
overestimated by $\sim0.2$ dex. The difference between the solar and
super-solar metallicity tracks is larger than for the FSPS models, but
this is mainly due to the fact that the BC03 super-solar metallicity
($Z=0.05$) track is significantly higher than the FSPS models
($Z=0.03$). For low-redshift galaxies in the SDSS, the super-solar
($Z=0.05$) model predicts a $M/L$ that is too low by $\sim 0.3$ dex,
while the scatter for the $z>0.5$ data with the super-solar model
(0.14) is almost the same as the scatter for the solar model
(0.15). The sub-solar (Z=0.004) model shows colors that are too blue
at fixed \mlg at all times.

In Figure \ref{fig:fig5}(d), there is a larger color separation between
the three metallicity tracks as compared to Figure \ref{fig:fig5}(c). At
fixed age the color difference in rest-frame $g-K$ is approximately
twice as large as the color difference in the rest-frame $g-z$ color,
indicating that the first color provides a better constraint for
metallicity (see also \citealt{bell2001}). Interestingly, the slope of
the BC03 models remain almost linear for the rest-frame $g-K$ color,
in contrast with the other two models. Neither solar nor super-solar
tracks provide a good match to all the data, but the super-solar
metallicity track has significantly lower scatter than the solar
metallicity (super-solar=0.20 vs. solar = 0.28).  While the solar
metallicity model is unable to reproduce the low $M/L$ of the bluest
galaxies, the super-solar metallicity model is unable to reproduce the
color and the $M/L$ for SDSS galaxies.
 			
\subsection{Ma11 models}
\label{subsec:ma11}
The Ma11 models are systematically different than the other models
(Figure \ref{fig:fig5}(e) and (f)). The Ma11 models exhibit an "S-shaped"
relation between the $M/L$ and rest-frame color, in particular for the
$g-K$ colors. The trend in the Ma11 models is such that for blue
colors there is a small increase in the $M/L$, whereas there is a
steep nearly vertical upturn in the $M/L$ at late ages. It is
interesting to note that the Ma11 solar metallicity track is able to
match the $M/L$ and rest-frame $g-z$ color of all
galaxies. Furthermore, the Ma11 model shows the lowest scatter as
compared to the other models: rms = 0.11.

However, for the \gk color the sub-solar and solar metallicity tracks
provide a poor match (Figure \ref{fig:fig5}(f)). The $Z=0.02$ model has
a sharp increase in the $M/L$ around \gk$\sim 1.4$, after which it is
too blue by $\sim0.5$ mag compared to the data. The super-solar
metallicity model is able to match all galaxies at $z<1$, but around 1
Gyr the $M/L$ is too low by $\sim0.2$ dex.

\begin{figure*}
\epsscale{1.05}
\plotone{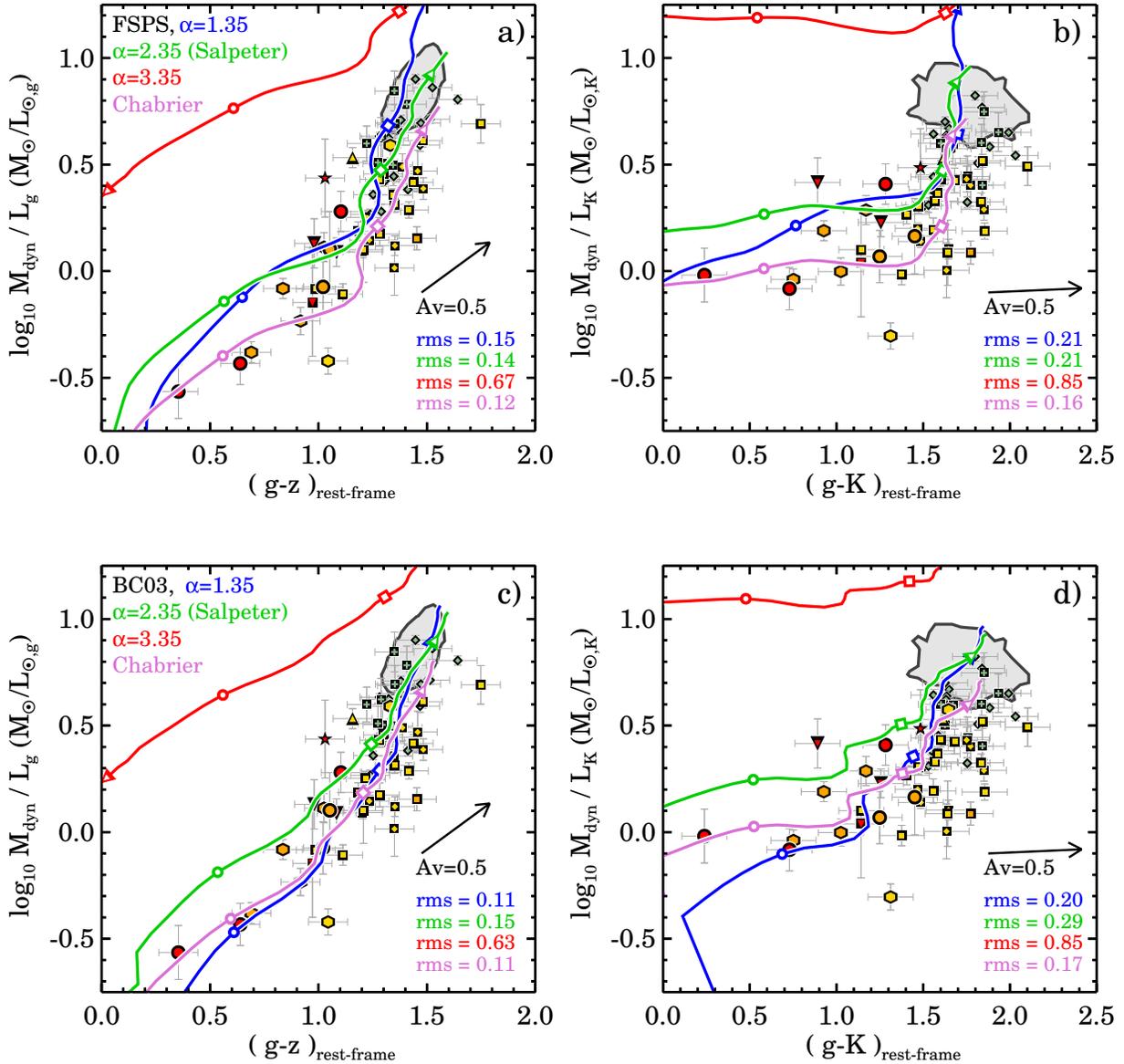}
\caption{$M/L$ vs. rest-frame color: comparing IMFs with different
slopes. We use FSPS (top row) and BC03 (bottom row) models with solar
metallicity ($Z = 0.02$). The IMF is defined as a single power-law
with slope $\alpha$. Different curves are IMFs with different
realizations of $\alpha$.  While the FSPS do not favor an IMF with
$\alpha =1.35$ (blue) over $\alpha=2.35$ (green), the BC03 has less scatter
for the IMF with $\alpha=1.35$ as compared to $\alpha=2.35$. Whereas a
Chabrier IMF is able to reproduce the low $M/L$ for the bluest
galaxies, it does not match the $M/L$ of low redshift SDSS galaxies in
Panel (a) and (c). It does shows an excellent match to the data in Panel
(b) and (d) with very little scatter.}
\label{fig:fig6}
\end{figure*}

\subsection{Summary}
\label{subsec:sum}
All the models reproduce the general observed trend in $M/L$
vs. rest-frame color.  Even though certain models with specific
metallicity tracks match one $M/L$ vs. color, none of the models are
able to simultaneously match the data in both the rest-frame $g-z$
vs. \mlg and $g-K$ vs. \mbox{\mlk.} In particular for FSPS and BC03,
the models are unable to match the low $M/L$ for the bluest galaxies
in combination with the rest of the data. For the \gz color, the Ma11
models are able to predict the low $M/L$ for the bluest galaxies at
the same time as the $M/L$ for galaxies in the SDSS, with lower rms
scatter as compared to the other models.  We furthermore find that the
SPS models exhibit different relations between the $M/L$ and
rest-frame color, most prominently visible in the rest-frame $g-K$
color.  The cause for the discrepancies between the models and the
data, but also among the different models, can be due to several
factors as there are a number of systematic differences in the SPS
models. It is beyond the scope of this paper to address these
differences, but we refer the reader to \citet{conroy2010} for a
recent comparison of several popular models.

\section{Constraints on the IMF}
\label{sec:imf}
In the previous Section we found that the models reproduce the general
observed trend between $M/L$ and color, but cannot match all the data. Adapting a different
IMF could provide a solution to this problem. The IMF influences the
evolution and scaling of the $M/L$, while it has a smaller effect on
the color evolution (\citealt{tinsley1972};
\citeyear{tinsley1980}). Generally speaking, a bottom-heavy IMF
($\alpha>2.35$) will give a flatter $M/L$ vs. color relation as
compared to a bottom-light IMF ($\alpha<2.35$). In this Section, we
explore the effect of the IMF on the different SPS models in the $M/L$
vs. color plane.

 						   		  						   		 
\begin{table}[t]
\caption{Scatter around SPS Models with different realizations of the IMF}
\label{tab:data4}
\begin{center}
\begin{tabular}{l l l| c c c c c}
\hline
\hline
 \colhead{SPS Model} & \colhead{$M/L$} & \colhead{color}   & \colhead{x=1.35} & \colhead{x=2.35} & \colhead{x=3.35}  & \colhead{Chabrier} \\
 \hline
FSPS     & $M/L_{\rm g}$ & $g-z$ &  0.15 & 0.14 & 0.67 & 0.12 \\
              & $M/L_{\rm K}$      & $g-K$ &  0.21 & 0.21 & 0.85 & 0.15 \\
                                    
BC03     & $M/L_{\rm g}$ & $g-z$ & 0.11 & 0.15 & 0.63 & 0.11 \\
               & $M/L_{\rm K}$     & $g-K$ & 0.20 & 0.29 & 0.85 & 0.17\\
\hline
\label {tab:tab6}
\end{tabular}
\end{center}
\end{table}


\begin{figure*}
\epsscale{1.05}
\plotone{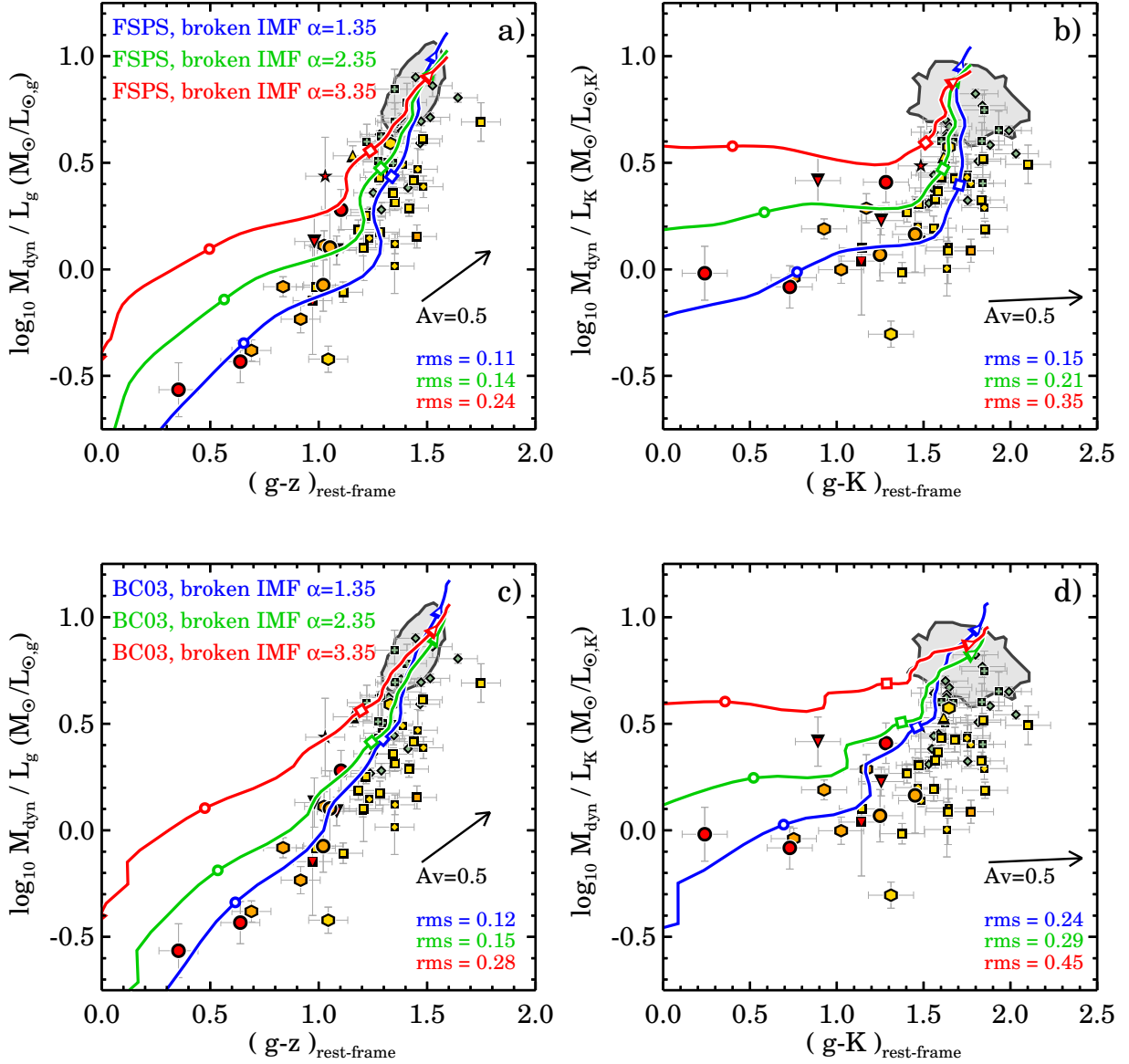}
\caption{$M/L$ vs. rest-frame color for an IMF with different
slopes between 1\msun and 4\msunp. Below 1\msun and above 4\msunp, this
IMF has a Salpeter slope, as defined according to Equation
\ref{eq:imf}. We use FSPS (top row) and BC03 (bottom row) models with
solar metallicity ($Z = 0.02$). Based on the rms scatter, the models
favor an IMF with a slope of $\alpha =1.35$ over $\alpha=2.35$. The
FSPS model with a broken IMF of $\alpha=1.35$ is able to reproduce
both the $M/L$ vs. $g-z$ and $g-K$ rest-frame color.}
\label{fig:fig7}
\end{figure*}


\subsection{IMF Comparison}
\label{subsec:imffit}

We show the FSPS (top row) and BC03 (bottom row) models with four
different realizations of the IMF in Figure \ref{fig:fig6}. We do not
further explore the Ma11 models, because these models with different
IMFs were not available to us. In this section we use solar
metallicity models (FSPS $Z = 0.0198$; BC03 $Z=0.02$), and a truncated
SFH with a constant star formation rate for the first 0.5 Gyr. The
Salpeter IMF with slope $\alpha=2.35$ is shown in green and was the
assumed IMF in Figure \ref{fig:fig5}. A bottom-light IMF with slope
$\alpha=1.35$ is shown in blue, the bottom-heavy $\alpha=3.35$ IMF in
red, and the Chabrier IMF in pink.

In Figure \ref{fig:fig6}(a) we find that the FSPS model with the
bottom-heavy IMF ($\alpha=3.35$) has a $M/L$ that is always too high
and does not match any of the data. The steepest $M/L$ vs. color
relation is predicted by the bottom-light IMF ($\alpha=1.35$). For the
Salpeter and the bottom-light IMF we measure a similar rms scatter
(0.14-0.15), but both IMFs have a $M/L$ that is on average too high by
$\sim0.1-0.2$ dex. The Chabrier IMF has a very similar behavior in the
$M/L$ vs. color plane as the Salpeter IMF, but with a lower $M/L$ by
about $\sim0.2$ dex. For the bluest galaxies the Chabrier IMF
reproduces the low $M/L$, but the $M/L$ is too low by $0.2$ dex for
galaxies in the SDSS. Out of all four realizations of the IMF that we
show in Figure \ref{fig:fig6}(a), we measure the least scatter for the
Chabrier IMF (0.12 dex).

For the \mlk vs. \gk (Figure \ref{fig:fig6}(b)), we find similar
differences between the IMFs as for the \mlk vs. \gz (Figure
\ref{fig:fig6}(a)). The bottom-heavy IMF ($\alpha=3.35$) overpredicts
the $M/L$ by more than a dex and does not match any of the
data. Compared to the Salpeter IMF, we find that the bottom-light IMF
has a steeper $M/L$ vs. color relation, with a steep vertical upturn
around 3 Gyr. Interestingly, the Chabrier IMF is able to match all
data with very little scatter (0.15 dex).


\begin{figure*}
\epsscale{1.05}
\plotone{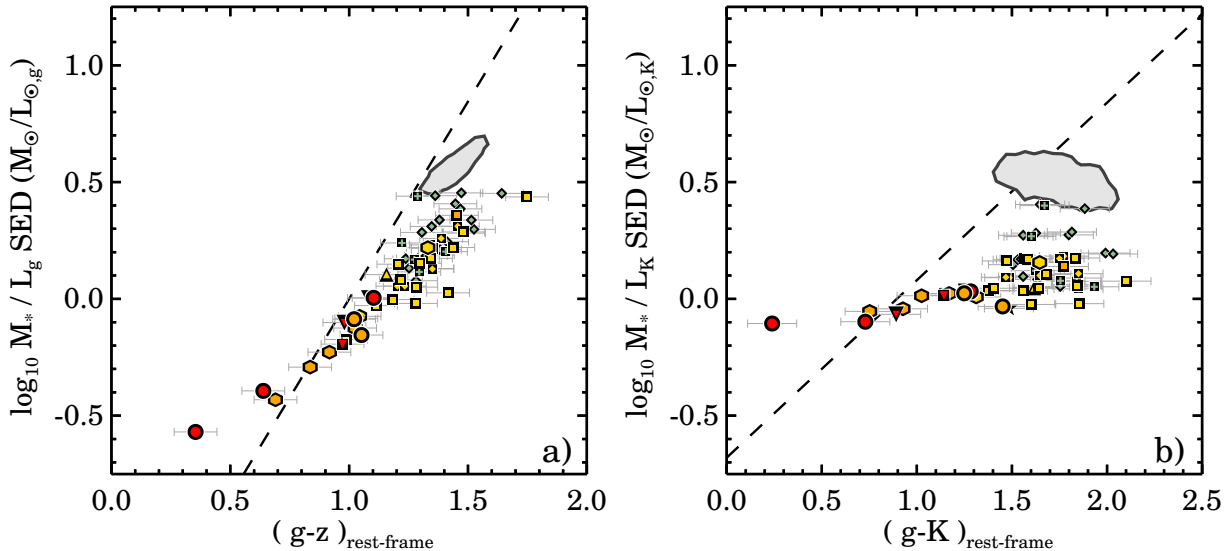}
\caption{$M_{*}/L_{\rm}$ versus rest-frame color. Here the $M/L$
have been determined by fitting solar metallicity BC03 models
to the full broad-band data. Panel a) The galaxies lie along a
tight sequence, but the relation is shallower as compared to the
best-fit dynamical relation (dashed-line) from Section
\ref{subsec:empfit}. Panel b) the SED derived $M_{*}/L_{\rm K}$
show a rather complex trend with \gk color similar to the trends for
the FSPS and Ma11 models.}
\label{fig:fig8}
\end{figure*}


In Figure \ref{fig:fig6}(c), we show the BC03 models with the four
different realizations of the IMF. Again, the bottom-heavy IMF does
not match any of the data. The bottom-light and Chabrier IMF both
provide an excellent match to the intermediate and high-redshift data
with the least rms scatter (0.11 dex). However, the Chabrier IMF again
predicts a $M/L$ that is slightly too low for the SDSS sample.

The bottom-light IMF which gave a perfect match for the \mlg vs. $\gz$,
however, does not provide a good match for the intermediate galaxies
in the \mlk vs. rest-frame $g-K$ color plane (Figure
\ref{fig:fig6}(d)). At $\gk>1.4$ the $M/L$ of the bottom-light IMF is on
average too high by $\sim0.2$ dex. The bottom-light IMF still provides
a better prediction than the Salpeter IMF with a respective rms of
0.20 vs 0.29. Again, we find that the Chabrier IMF best matches all
the data based on the rms scatter (0.17), but does not provide a
perfect match either.

 						   		  						   		 
\begin{table}[t]
\caption{Scatter around SPS Models with different realizations of the broken IMF}
\label{tab:data5}
\begin{center}
\begin{tabular}{l l l | c c c c c}
\hline
\hline
 \colhead{SPS Model} & \colhead{$M/L$} & \colhead{color}   & \colhead{x=1.35} & \colhead{x=2.35} & \colhead{x=3.35}  \\
 \hline
FSPS     & $M/L_{\rm g}$           & $g-z$ & 0.11 & 0.14 & 0.24  \\
        & $M/L_{\rm K}$            & $g-K$ & 0.15 & 0.21 & 0.35 \\
                                    
BC03     & $M/L_{\rm g}$             & $g-z$ & 0.12 & 0.15 & 0.28 \\
         & $M/L_{\rm K}$           & $g-K$ & 0.24 & 0.29 & 0.45 \\
\hline
\label{tab:tab7}
\end{tabular}
\end{center}
\end{table}

Overall, we find that the BC03 model favors an IMF with a slope of
$\alpha=1.35$ over an IMF with $\alpha=2.35$, while for the FSPS
models we find no statistical preference for one of the two. We get an
excellent match to the data with the FSPS model when using a Chabrier
IMF for the rest-frame $g-K$ color, but for the rest-frame $g-z$ color
this IMF underpredicts the $M/L$ for galaxies in the SDSS. As we still
have not identified a model that can simultaneously match the $M/L$
vs. \gz and \gk colors, we explore a more exotic IMF in the next
Section.

\subsection{Broken IMF}
\label{subsec:brokenimffit}
As the $M/L$ vs. color relation is mostly sensitive to the IMF around
the main sequence turnoff-point of stars in the Hertzsprung-Russell diagram, we experiment with a
broken IMF, in which we only vary the slope between 1\msun and 4\msunp:
\begin{eqnarray}
 \frac{dN}{dM} & \propto  M^{-2.35} &\quad\text{for,~~~}  [0.08 < M_* / M_{\odot} < 1] \\
 \label{eq:imf}
 \frac{dN}{dM} &\propto  M^{-\alpha} &\quad\text{for,~~~} [1 < M_* / M_{\odot} < 4] \\
 \frac{dN}{dM} &\propto  M^{-2.35}  &\quad\text{for,~~~} [4 < M_* / M_{\odot} < 100]
\end{eqnarray}
One advantage of this approach is that different realizations of the
IMF will cause the SPS tracks to naturally intersect at late ages when
most of the integrated light will come from low-mass stars with $M_*<
1$\msunp.

Figure \ref{fig:fig7} shows the three different realizations of the
IMF using Equation \ref{eq:imf}: $\alpha=1.35$ (blue), $\alpha=2.35$
(green, normal Salpeter), and $\alpha=3.35$ (red). As before, we use
solar metallicity models (FSPS $Z = 0.0198$; BC03 $Z=0.02$), and a
truncated SFH with a constant star formation rate for the first 0.5
Gyr.

As expected, in Figure \ref{fig:fig7}(a) we find that the different
tracks now all match the oldest $z\sim0$ SDSS galaxies. This Figure
also clearly shows that the $M/L$ vs. color relation becomes
increasingly steep with decreasing slope of the IMF. We find that the
FSPS model with $\alpha=1.35$ IMF is able to reproduce the low $M/L$
for the bluest galaxies and matches all the other data as well, with
very little scatter (rms=0.11). Furthermore, the broken IMF with
$\alpha=1.35$ provides a better match to the data than the IMF with
slope $\alpha=2.35$. The bottom-heavy $\alpha=3.35$ IMF matches the highest $M/L$
galaxies, but for all galaxies bluer than \gz$<1.2$ the model $M/L$ is
still too high. Most interestingly, in Figure \ref{fig:fig7}(b) the FSPS
model with $\alpha=1.35$ IMF matches all the data for the \mlk vs \gk
with very little scatter (rms=0.15).

In Figure \ref{fig:fig7}(c) we show the BC03 models with different
realizations of the broken IMF. The broken IMF with $\alpha=1.35$
matches all the data from $z\sim2$ to $z\sim0$. However, for the \mlk
vs. \gk color this IMF over-predicts the $M/L$ by $\sim0.2$ dex
(Figure \ref{fig:fig7}(d)).

Therefore, based on the rms scatter, we conclude that both the FSPS
and BC03 models favor a slope of the broken IMF of $\alpha=1.35$ over
$\alpha=2.35$. We note that the FSPS model with a broken IMF of
$\alpha=1.35$ is the only model that can reproduce both the $M/L$
vs. $g-z$ and $g-K$ rest-frame color.

\section{Discussion and Comparison to Previous Studies}
\label{sec:disc}
\subsection{SED derived M/L}
\label{subsec:sedml}

To investigate the implication of the results, we first compare our
relation of $M_{\rm dyn}/L$ vs. rest-frame color to the relation of
$(M_*/L)_{\rm SED}$ vs. rest-frame color. The $(M_*/L)_{\rm SED}$ has
been determined by fitting solar metallicity BC03 models to the full
photometric broad-band dataset (see Section \ref{subsec:galprop}). We
show the results for our sample in Figure \ref{fig:fig8}. In Figure
\ref{fig:fig8}(a) we compare the $(M_*/L_{\rm g})_{\rm SED}$
vs. rest-frame $g-z$ color with our best-fit dynamical relation from
Section {\ref{subsec:empfit} (dashed line). As expected, the galaxies
lie along a tight sequence. We note, however, that the best-fit
dynamical relation is steeper. This difference in the steepness of the
relation is consistent with the fact that the BC03 model tracks do not
quite track the trends of Figure \ref{fig:fig5}c. Figure
\ref{fig:fig8}(b) shows the $(M_*/L_{\rm K})_{\rm SED}$ versus $\gk$.
The derived $(M_*/L_{\rm K})_{\rm SED}$ show a rather complex trend
with \gk color. This complex trend is similar to the trends for the
FSPS (Figure \ref{fig:fig5}(b)) and Ma11 (Figure \ref{fig:fig5}(f)) SPS
models.

The mismatch between the $(M_*/L_{\rm g})_{\rm SED}$ (using the BC03
models which were used to fit the full photometric broad-band dataset) 
and the \mlg is highlighted in Figure \ref{fig:fig9}, where we
compare the two estimates directly. In the case that the dynamical
$M/L$ corresponds well to the SED based $M/L$, we expect to see a
one-to-one linear relation (dashed line) with potentially a constant
offset due to dark matter or low-mass stars. However, Figure
\ref{fig:fig9} shows a relation that has a shallower slope than the
dashed-line, such that galaxies with a lower $M/L$ are further offset
from the one-to-one relation. This non-constant offset is similar to
the results by \citet{vandesande2013}, in which we showed that
$M_{*}/M_{\rm dyn}$ changes slightly as a function of redshift, where
the $z\sim2$ galaxies had the highest $M_{*}/M_{\rm dyn}$. However, as
redshift, color, and $M_{\rm dyn}/L$ are correlated in our sample (see
Figure \ref{fig:fig2}(a) and (c)), the $M_{*}/M_{\rm dyn}$ trend with
redshift could also be caused by the non-constant offset in
$(M_*/L_{\rm g}) _{\rm SED}$ vs. $\mlg$. Without additional
information (e.g., high signal-to-noise spectroscopy) it is hard to
establish whether the trend is driven by galaxy structure
evolution/dark matter fraction evolution, IMF variations (Section
\ref{sec:imf}), or discrepancies in SPS models.

\subsection{Intrinsic scatter}
From the \mlg vs \gz color relation, we find that we can predict the
$M/L$ of a galaxy with an accuracy of $\sim0.25$ dex. However, our
dynamical $M/L$ estimates suffer from large (systematic)
uncertainties.  To quantify the intrinsic scatter in the relation, we
calculate the fraction of the scatter induced by uncertainties in the
size, and velocity dispersion measurements. As the formal errors on
the rest-frame colors are small ($<0.01$ mag), the scatter will be
dominated by errors in the dynamical mass. From Monte-Carlo
simulations, we find that 0.11 dex of the 0.25 dex scatter can be
explained by measurement uncertainties. Thus our intrinsic scatter in
the \mlg vs. \gz relation is 0.22 dex. However, if the (systematic)
errors on the rest-frame colors are larger, for example $\sim0.1$ mag,
the intrinsic scatter would be 0.14 dex. Furthermore, from the Bayesian linear fit we
obtain an estimate for the intrinsic scatter of 0.15 dex for the \gz vs. $\mlg$. 
The intrinsic scatter can be due to variations in the star formation histories, 
metallicities for galaxies at high and low redshift, or
may be due to unknown sources of measurement errors.

We also consider the fact that a single color might not provide
an accurate constraint and that the full broad-band SED fit yields a
tighter relation. We use the SED derived stellar $M/L$ (see also
Section \ref{subsec:sedml}) to estimate the scatter when using the
full broad-band dataset. We find a mean ratio of $M_{*} / M_{\rm dyn}
= -0.20$ with an rms scatter of 0.20 dex. This scatter of 0.20 dex
suggests that a single color (which had a $M/L$ accuracy of 0.25 dex)
only provides a slightly worse $M/L$ prediction as compared to full
broad-band SED fitting.

\subsection{Comparison to Literature}
\subsubsection{Single-burst SPS models}
\label{subsec:compspslit}

The evolution of the rest-frame K-band $M/L$ out to $z\sim1$ was
measured for the first time by \citet{vanderwel2006}, for a sample
with a small dynamic range of approximately $\sim 0.4$ dex in
$M/L_{\rm K}$. They concluded that single-burst BC03 models with a
Salpeter IMF were offset with respect to the data and that the Ma05
models with a Salpeter IMF provided the best match. While we come to a
similar conclusion for the BC03 models, we still find a large offset
between the data and the Maraston models with a Salpeter IMF, as the
Maraston solar metallicity tracks are too blue by 0.5 mag in
rest-frame $g-K$. The different conclusion can be explained by the
fact that \citet{vanderwel2006} used relative $M/L$ and colors, while
we only use absolute values. We could not find other direct
comparisons between $M_{\rm dyn}/L$ vs. color relations and SPS models
in the literature.

\begin{figure}
\epsscale{1.05}
\plotone{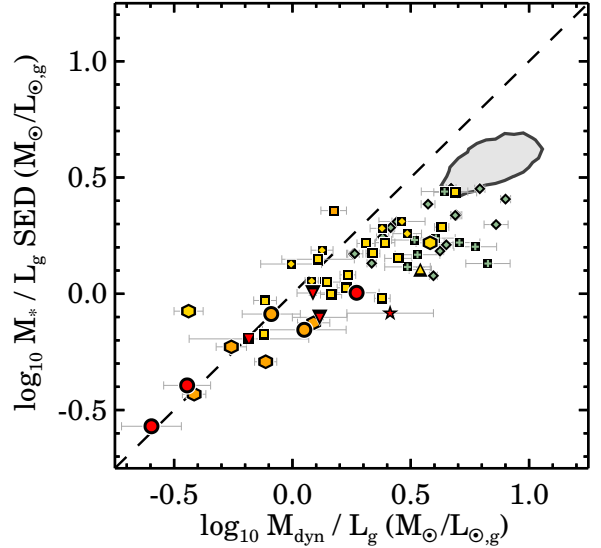}
\caption{$M_{*}/L_{\rm g}$ from SED fits vs. the $\mlg$. The $M/L$
have been determined by fitting solar metallicity BC03 models
to the full broad-band data.  We find a non-linear relation, which
could be due to an evolving dark matter fraction or IMF variations.}
\label{fig:fig9}
\end{figure}


\begin{figure*}
\epsscale{1.05}
\plotone{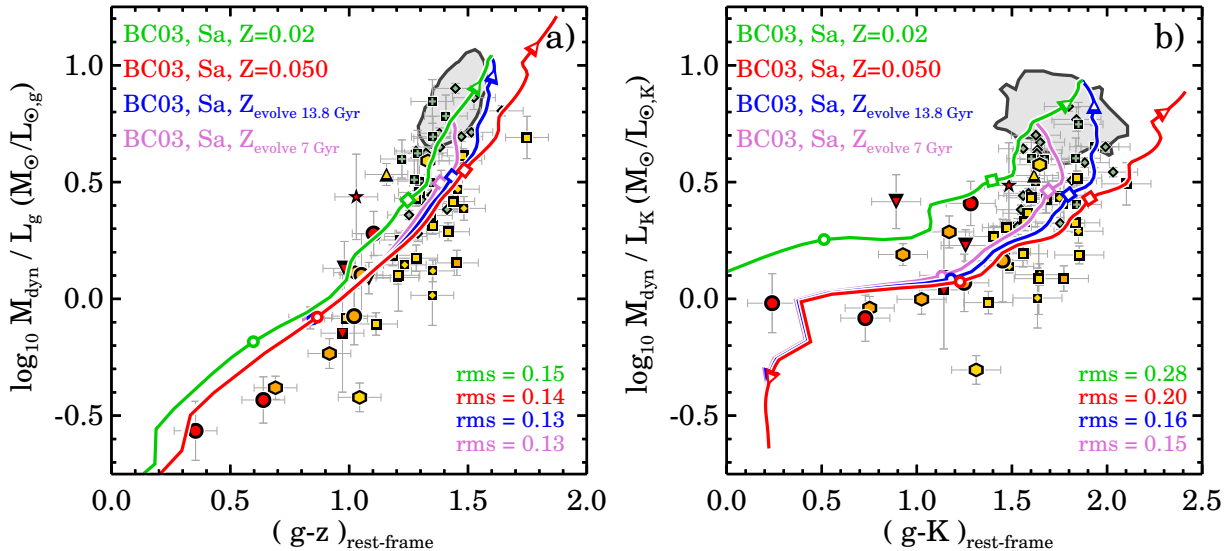}
\caption{$M/L$ vs. rest-frame color for evolving metallicities with BC03 models. 
The green ($Z=0.02$ and red ($Z=0.05$) model track are the same as in Figure 
\ref{fig:fig5}(c) and (d). The blue track is a model where the metallicity evolves from
$Z=0.05$ to $Z=0.02$ a function of time, over a period of 13.8 Gyr. The pink 
is similar to the blue track, but for this model the metallicity evolves from from 
$Z=0.05$ to $Z=0.02$ within the first 7 Gyr. The models with evolving metallicity 
show a slightly steeper relation as compared to the models without evolution in 
the metallicity.}
\label{fig:fig10}
\end{figure*}


\subsubsection{Star-Forming Galaxies and extended SFH}
\label{subsec:litmlvscol}

The relation between $M/L$ vs. color was first explored by
\citet{bell2001} for spiral galaxies. They used an early version of
the BC03 models, and found that the rest-frame $B-R$ color provided a good estimate of
the $M/L$. Follow up work by \citet{bell2003} used the Pegase SPS
models with more extended SFHs to estimate SED based $M/L$, which were
then used to derive observationally-constrained $M/L$ vs. color
relations. They found that the optical $M/L$ vs. color relations were
in good agreement with \citet{bell2001}, but they found a shallower
slope in the NIR $M/L$ vs. color relation due to unaccounted
metallicity effects. \citet{zibetti2009} used the latest SPS models
from Charlot \& Bruzual (in Prep.) to directly derive the $M/L$
vs. color relation. Similar to the results by
\citet{gallazzi2009} they found a steeper slope in the $M/L_{\rm
i}$ vs. $(g-i)_{\rm rest-frame}$ relation as compared to
\citet{bell2003}. Using BC03 models \citet{taylor2011} follow a
similar method as \citet{bell2003} and found that slope for the
$M/L_{\rm i}$ vs. $(g-i)$ relation is steeper than in \citet{bell2003},
but shallower than in \citet{zibetti2009}. \citet{into2013} use the latest Padova
isochrones, with detailed modelling of the Thermally Pulsing Asymptotic Giant Branch
phase, to update the theoretical $M/L$-color relations. They also find a steeper slope 
for their new relations as compared to \citet{bell2003}.

In this Paper, we found that the slope for the $M/L_{\rm i}$
vs. rest-frame $(g-i)$ relation is considerably steeper than in
previous work (e.g, \citealt{bell2003}; \citealt{zibetti2009};
\citealt{taylor2011}; \citealt{into2013}). The difference is easily explained as the $M/L$
vs. color relations in previous studies were derived from samples that
include star-forming galaxies, with variable (exponentially declining)
star formation histories, and dust. This naturally leads to a
shallower slope. In addition, we found in Section \ref{sec:comp} that
most model tracks predict a shallower relation as compared to our
dynamical data. If this trend is indeed caused by stellar population
effects, it would imply that the masses of star forming galaxies need
recalibration, and may have systematic uncertainties at a level of 0.2
dex.

\subsection{Constraints on the IMF}
\label{subsec:disimf}

Several authors have constrained the IMF's using a differential
analysis of color evolution against $M/L$ evolution, inspired by early
work by \citeauthor{tinsley1972} (\citeyear{tinsley1972};
\citeyear{tinsley1980}). \citet{vandokkum2008a} used galaxies in
clusters at $0< z< 0.8$ and found an IMF slope around $1M_{\odot}$ of
$\alpha = 0.7^{+0.4}_{-0.7}$. \citet{holden2010} used a larger sample,
and analyzed the evolution at fixed velocity dispersion, and found an
IMF slope of $\alpha=1.9\pm0.2$.  \citet{vandokkum2012} repeated this
analysis with the latest population models \citep{conroy2012} and
found a slope of $\alpha = 1.81\pm0.27$. Although the techniques used
by these authors are quite different, the results are similar to
those presented here. However, with our larger range in $M/L$ and
rest-frame color, we find that the variations between different models
with the same IMF are comparable to the variations due to the IMF with
the same model. Therefore, with the current models it is still hard to
put a robust constraint on the IMF.

\subsection{Dark Matter}
\label{subsec:dm}

In this Paper we use dynamical mass estimates for calculating the
$M/L$.  The dynamical mass includes both stellar mass and dark matter
mass, but to this point we have ignored the contribution of dark
matter to the dynamical mass. At low-redshift the dynamical to stellar
mass fraction is approximately a factor of $~1.6$ within one effective
radius, due to the contribution of dark matter to the total mass. If
we include dark matter in the $M/L$ of the models, this fraction would
shift all curves in Figure \ref{fig:fig5}-\ref{fig:fig7} vertically up
by $\sim0.2$ dex. This shift would not solve the discrepancies between
the models and the data, because the discrepancies are in the slope
and cannot be solved by a constant offset (see Figure \ref{fig:fig9}).

However, whether the dark matter fraction within one $r_{\rm e}$ is
constant over time is still subject to debate.  The size growth of
massive quiescent galaxies may result in an increase of the dark
matter to stellar mass fraction within one $r_{\rm e}$, because the
dark matter profile is less steep than the stellar mass profile (see
also \citealt{hilz2013}). Thus, the dark matter fraction within one
$r_{\rm e}$ may increase over time. In \citet{vandesande2013}, we
indeed find a hint of an evolving dark matter fraction, i.e, the
median $M_{*}/M_{dyn}$ is higher by $50\%$ at $z>1.5$ compared to
massive SDSS galaxies ($M_{*}/M_{dyn} \propto (1+z)^{0.17\pm
0.011}$). From hydrodynamical simulations, \citet{hopkins2009a} find
that for galaxies with $M_* \sim 10^{11}$, the stellar to dynamical
mass ($M_{*}/M_{dyn}$) at $z\sim2$ is lower by 0.1 dex.  Thus, if we
correct for an evolving dark matter fraction the $M/L$ for
high-redshift galaxies would decrease by approximately $0.1$ dex and
for SDSS galaxies by about $0.2$ dex.  As this correction decreases
the slope of the empirical $M/L$ vs. color relation, it would make the
slope of the data more consistent with a Chabrier ($\alpha=2.35$) IMF 
(see Figure \ref{fig:fig6}).

\subsection{Metallicity \& Complex Star Formation Histories}
\label{subsec:metal}

In the comparison of the models with the data, we used model tracks
with single metallicities. As galaxies grow in size and mass over
time, for example through minor mergers, metallicity may also evolve
as the satellite galaxies have lower metallicities (e.g.,
\citealt{gallazzi2005}; \citeyear{gallazzi2014};
\citealt{choi2014}). The core will likely keep the same metallicity,
while the metallicity in the outskirts may decrease
(\citealt{greene2013}; \citealt{montes2014}). 
Also, due to our stellar mass selection limit of $10^{11}$\msunp, 
the descendent of the $z=2$ galaxies will be more massive than our $z=0$
selected galaxies. Thus, the average metallicity of $z=0$ galaxies in our 
sample is likely lower than that of the $z=2$ galaxies.

We make two simple models for which the metallicity is allowed to evolve as 
function of time from supersolar ($Z=0.05$) to solar ($Z=0.02$). In Figure 
\ref{fig:fig10} we show the BC03 models, similar to Figure \ref{fig:fig5}(c) 
and (d). In blue, we show the model track with a metallicity transition timescale 
of 13.8 Gyr. For the pink model track, we assume that the metallicity evolution 
occurred within the first 7 Gyr after the burst. In Figure \ref{fig:fig10}(a) the 
metallicity evolution has very little effect as the $Z=0.02$ and $Z=0.05$ 
metallicity tracks are close together. The rms scatter is only slightly lower 
for the model tracks with metallicity evolution (rms = 0.13) as compared to 
normal metallicity tracks (rms = 0.14-0.15). In Figure \ref{fig:fig10}(b), 
both metallicity evolution tracks (blue and pink) show a steep vertical upturn 
in the $M/L$ around 3 Gyr, which improves the match with the observed data. 
The rms scatter is significantly lower for the models where the metallicity is 
allowed to evolve (e.g., 0.15 vs. 0.20).

While the BC03 models with metallicity evolution provide a better match for 
the \mlk vs. \gk (Figure \ref{fig:fig10}(b)), these models do not provide a 
significant improvement for the \mlg vs. \gz, in particular at blue colors 
and low $M/L$. Without additional data (e.g., resolved images and spectroscopy) 
we cannot further quantify the effect of metallicity evolution.

Finally, we have assumed a single SFH for all galaxies. While
massive galaxies in general are thought to have simple SFHs, for
individual galaxies the SFH could be far more complex due to merging
events. The fact that we find that none of the SPS models with a Salpeter or Chabrier IMF are able to
simultaneously match all the data for both the rest-frame optical and
NIR data could imply that the effect of a complex SFH is more
important than assumed here.
 
\subsection{Systematic Sample Variations}
\label{subsec:sys}

While the approach of using a mass selected sample has provided us
with many insights it is clear that for comparing the $M/L$ of
galaxies at different redshifts, this static mass selection could
introduce a bias. Recent studies find that several properties of
massive quiescent galaxies may change over time: they were smaller
than their present-day counterparts (e.g., \citealt{daddi2005};
\citealt{trujillo2006}; \citealt{vandokkum2008b}; \citealt{franx2008};
\citealt{vanderwel2008b}; and numerous others), their stellar masses
increase by a factor of $\sim2$ from $z\sim2$ to $z\sim0$ (e.g.,
\citealt{vandokkum2010a}, \citealt{patel2013}), and the effective
velocity dispersion may also decrease (e.g., \citealt{oser2012};
\citealt{vandesande2013}). Thus, our samples and measurements at
different redshifts may not be directly comparable.

A possible additional complication is progenitor bias (e.g.,
\citealt{vandokkum1996}; \citeyear{vandokkum2001}): the number density
of massive galaxies changes by a factor of $\sim10$ from $z\sim2$ to
$z\sim0$ \citep{muzzin2013b}. Thus, a substantial fraction of the
current day early-type galaxies were star-forming galaxies at $z\sim2$
(see also \citealt{vanderwel2009}). If the properties of the
descendants of these $z\sim2$ star-forming galaxies are systematically
different from the descendants of the quiescent galaxies at $z\sim2$,
the simple single burst SPS models which we used here may produce a
biased result.

%

%
\section{CONCLUSIONS}
\label{sec:conclusions}

In this Paper, we have used a sample of massive galaxies ($M_*>10^{11}
M_{\odot}$) out to $z\sim2$ with stellar kinematic, structural, and
photometric measurements. The primary goals of this Paper are to study
the empirical relation between the dynamical $M/L$ and rest-frame
color, assess the ability of SPS models to reproduce this relation,
and study the effect of the IMF on the $M/L$ vs color relation.

We find that our sample spans a large range in $M/L$: 1.8 dex in
rest-frame $\log M_{\rm{dyn}}/L_{\rm u}$, 1.6 dex in $\log
M_{\rm{dyn}}/L_{\rm g}$, and 1.3 dex in $\log M_{\rm{dyn}}/L_{\rm
K}$. As expected for a passively evolving stellar population, we find
a strong correlation between the $M/L$ for different bands and
rest-frame colors. For rest-frame optical colors, the correlation is
well approximated by a linear relation, and we provide coefficients of
the linear fits for a large number of $M/L$ vs. color
correlations. The root-mean-square scatter in the $\log
M_{\rm{dyn}}/L$ residuals is $\sim0.25$ dex.  Thus, these relations
are ideal for estimating masses for quiescent galaxies with an
accuracy of $\sim0.25$ dex.

We compare a combination of two $M/L$ vs. rest-frame color relations
with SPS models by \citet{bruzual2003},
\citet{maraston2011}, and \citet{conroy2009}. Under the assumption of
a Salpeter initial mass function (IMF), none of the SPS models are
able to simultaneously match the data in $M_{\rm{dyn}}/L_{\rm g}$
vs. $(g-z)_{\rm rest-frame}$ color and $M_{\rm{dyn}}/L_{\rm K}$
vs. $(g-K)_{\rm rest-frame}$ color.

By changing the IMF, we test whether we can obtain a better match between the models and the data.
IMFs with different slopes are still unable to simultaneously match
the low $M/L$ of the bluest galaxies in combination with the other
data. While a Chabrier IMF underpredicts the $M/L$ for $z\sim0$ SDSS
galaxies in the $M_{\rm{dyn}}/L_{\rm g}$ vs. $(g-z)_{\rm rest-frame}$,
it provides an excellent match to all other data.

We also explore a broken IMF with a Salpeter slope at \mbox{$M<1$\msun}
and \mbox{$M>4$\msunp}, and we find that the models favor a slope of
$\alpha=1.35$ over $\alpha=2.35$ in the intermediate region, based on
the rms scatter. This time, the FSPS solar metallicity model with an IMF slope of
$\alpha=1.35$, is able to simultaneously match both the $M/L$ vs. \gz
and $\gk$ relations.

The combination of the $M/L$ and color is a powerful tool for studying
the shape of the IMF near 1\msun. However, this work shows that
the variations between different SPS models are comparable to the
variations induced by changing the IMF.  There are several caveats
which may change our data or models tracks, among which an evolving
dark matter fraction, an evolving metallicity, complicated star
formation histories, and an evolving mass-selection limit. More
complete and higher resolution empirical stellar libraries, improved
stellar evolution models, and larger spectroscopic samples at
high-redshift, are needed to provide more accurate constraints on the
IMF.

%
\vspace{0.2cm}

\acknowledgments{
We thank the anonymous referee for the constructive comments 
which improved the quality and readability of the paper.  
It is a pleasure to acknowledge the contribution to this work by the
NMBS and 3DHST collaboration. We also thank Rik Williams and Ryan
Quadri for their help with the UDS catalogs, and thank Andrew Newman
for providing the corrected stellar masses. The authors furthermore
wish to thank Daniel Szomoru and Adam Muzzin interesting discussions
which contributed to this Paper.  This research was supported by
grants from the Netherlands Foundation for Research (NWO), the Leids
Kerkhoven-Bosscha Fonds. This work is based on observations taken by
the 3D-HST Treasury Program (GO 12177 and 12328) with the NASA/ESA
HST, which is operated by the Association of Universities for Research
in Astronomy, Inc., under NASA contract NAS5-26555.}

\newpage


\clearpage

\end{document}